# Characterizing Protoclusters and Protogroups at $z \sim 2.5$ Using Ly$\alpha$ Tomography

Mahdi Qezlou[1,2,3] , Andrew B. Newman[2] , Gwen C. Rudie[2] , and Simeon Bird[1]

[1] Department of Physics & Astronomy, University of California, Riverside, CA 92521, USA; mahdi.qezlou@email.ucr.edu
[2] The Observatories of the Carnegie Institution for Science, 813 Santa Barbara Street, Pasadena, CA 91101, USA

*Received 2021 December 7; revised 2022 March 17; accepted 2022 March 28; published 2022 May 10*

## Abstract

Ly$\alpha$ tomography surveys have begun to produce 3D maps of the intergalactic medium opacity at $z \sim 2.5$ with megaparsec resolution. These surveys provide an exciting new way to discover and characterize high-redshift overdensities, including the progenitors of today's massive groups and clusters of galaxies, known as protogroups and protoclusters. We use the IllustrisTNG-300 hydrodynamical simulation to build mock maps that realistically mimic those observed in the Ly$\alpha$ Tomographic IMACS Survey. We introduce a novel method for delineating the boundaries of structures detected in 3D Ly$\alpha$ flux maps by applying the watershed algorithm. We provide estimators for the dark matter masses of these structures (at $z \sim 2.5$), their descendant halo masses (at $z = 0$), and the corresponding uncertainties. We also investigate the completeness of this method for the detection of protogroups and protoclusters. Compared to earlier work, we apply and characterize our method over a wider mass range that extends to massive protogroups. We also assess the widely used fluctuating Gunn–Peterson approximation applied to dark-matter-only simulations; we conclude that while it is adequate for estimating the Ly$\alpha$ absorption signal from moderate-to-massive protoclusters ($\gtrsim 10^{14.2} \ h^{-1} M_\odot$), it artificially merges a minority of lower-mass structures with more massive neighbors. Our methods will be applied to current and future Ly$\alpha$ tomography surveys to create catalogs of overdensities and study environment-dependent galactic evolution in the Cosmic Noon era.

*Unified Astronomy Thesaurus concepts:* Dark matter distribution (356); Large-scale structure of the universe (902); Galaxy environments (2029); Protoclusters (1297); Lyman alpha forest (980); Intergalactic medium (813); Dark matter density (354); Magnetohydrodynamical simulations (1966)

## 1. Introduction

The ancestors of massive galaxy groups and clusters are already evident as unvirialized large-scale structures at "cosmic noon," redshifts $z \sim 2 - 3$. These "protoclusters" and "protogroups" are moderate matter overdensities of $\delta_m = \rho / \bar{\rho} - 1 \approx 2 - 4$ that span $\sim 5 - 10 \ h^{-1}$cMpc (Chiang et al. 2013) and form the nodes of the cosmic web. Locating and characterizing these overdensities is essential to understanding the onset of the relation between galaxy properties and large-scale environments. In the local universe, observations have established a correlation between galaxy properties, such as star formation and galaxy morphologies, and the local density (Dressler 1980; Postman & Geller 1984; Kauffmann et al. 2004; Peng et al. 2010). Some recent studies suggest that these trends may persist to higher redshifts, $z \gtrsim 2.5$ or even $z \approx 3.5$ (Chartab et al. 2020; Lemaux et al. 2020). However, the current sample of known protoclusters is small and may suffer from important biases; resolving them requires new methods (for a review, see Overzier 2016).

Unlike clusters at $z \lesssim 1.5$, protoclusters at $z \sim 2.5$ are unvirialized structures and have not yet developed sufficiently hot gas to be observed via the Sunyaev–Zel'dovich effect or X-ray observations. Thus the traditional method for detecting protoclusters and groups at $z > 2$ is instead to search for overdensities of galaxies. However, using galaxies to locate protoclusters has some disadvantages. Most methods use a single type of galaxy to trace overdensities: sometimes typical star-forming galaxies, but often rare subtypes such as radio galaxies or extremely infrared-luminous systems (Miley & De Breuck 2008; Planck Collaboration et al. 2015; Noirot et al. 2018). While such methods successfully identify overdensities, studying how galactic evolution is affected by density is complicated when the galaxies used to locate overdense structures are required to have particular properties. Moreover, with current facilities, dense spectroscopic sampling of cosmological volumes is very challenging, while photometric surveys suffer from imprecise redshift estimates.

An exciting and complementary method for locating protoclusters and protogroups is Ly$\alpha$ absorption, which is particularly advantageous as it does not depend on galaxies to trace the underlying density field. Rather, it relies on neutral hydrogen gas, which traces the underlying density on large scales with high fidelity (Rauch 1998; McQuinn 2016). Ly$\alpha$ absorption has conventionally been observed in the spectra of bright background quasars and has enabled the detection of the H I reservoirs associated with several large-scale structures (e.g., Cai et al. 2016, 2017).

However, protoclusters have a complex 3D structure that cannot be fully characterized using 1D information from the spectra of one (or a few) bright quasars. Earlier theoretical works developed methods to optimally reconstruct the 3D structure of the intergalactic medium (IGM) from a set of spectra of closely spaced background sources (Pichon et al. 2001), a technique known as Ly$\alpha$ (or IGM) tomography. By testing against numerical simulations, they proved these methods recover the cosmic web structure with a high fidelity (Caucci et al. 2008; Lee & White 2016). On the observational side, Lee et al. (2014) showed that faint star-forming galaxies (Lyman-break galaxies, or LBGs) are suitable background sources due to their much higher abundance compared to QSOs. The current generation of telescopes is already sensitive enough to observe the Lyα forest absorption within a large sample of these faint LBGs, sufficient to map the IGM in 3D with a resolution of few $h^{-1}$cMpc.

---

[3] UCR-Carnegie Graduate Fellow.

To date, there are only two tomography surveys dedicated to mapping the IGM with megaparsec resolution at $z \sim 2.5$. The COSMOS Lyα Mapping and Tomography (CLAMATO) project (Lee et al. 2018; Horowitz et al. 2021a), the pioneering survey of this type, created the first megaparsec-resolution large-scale structure map in the redshift range $z = 2.05 - 2.55$ comprising a survey volume of $4.1 \times 10^5 \, h^{-3}$ Mpc³. This map led to the successful detection of a protocluster with an estimated mass of $9 \times 10^{13} \, h^{-1} M_\odot$ at $z \sim 2.45$ within the COSMOS field (Lee et al. 2016) along with a characterization of voids at the same redshift (Krolewski et al. 2018). These studies demonstrated the capacity of 3D Lyα tomography as a systematic way to detect and study large-scale structures. A larger experiment, the Lyα Tomography IMACS Survey (LATIS), is underway at the Magellan Baade telescope (Newman et al. 2020). LATIS achieves a similar, but slightly lower, sightline density to CLAMATO over a volume ≈10× larger, sufficient to map a statistically representative sample of large-scale overdensities. Newman et al. (2020) detected 36 significant over- or underdense regions within the redshift range of $z = 2.2 - 2.8$ using the first one-third of observations.

In this work, we are interested in developing methods to characterize the large set of structures that will be cataloged in the LATIS survey; however, our methods apply to other similar surveys. Using mock observations generated from cosmological hydrodynamical simulations, we improve on existing methods to detect and estimate the masses of structures detected in Lyα flux maps. Previous works have mainly focused on characterizing the most prominent structures, i.e., protoclusters (Stark et al. 2015b; Lee et al. 2016). However, many significant overdensities will collapse into slightly less massive halos by $z = 0$, i.e., protogroups. We define protoclusters and protogroups to be the progenitors of $z = 0$ halos with masses $M_{vir}/(h^{-1} M_\odot) > 10^{14}$ and $10^{14} > M_{vir}/(h^{-1} M_\odot) > 10^{13.5}$, respectively. To maximize the utility of current surveys' tomographic maps, we extend earlier methods to enable the detection and characterization of the progenitors over a wider range of mass. We also introduce a new method to partition large structures composed of multiple substructures. This delineation is of great value for galaxy–environment relation studies.

To achieve a volume representative of the ongoing and near-future surveys, simulations of size over a few 100 $h^{-1}$cMpc are required. Previous works mainly employed dark-matter-only simulations to model the signal in the tomography maps and to tune models for detecting protoclusters (Stark et al. 2015b; Lee et al. 2016) or voids (Stark et al. 2015a; Krolewski et al. 2018; Newman et al. 2020). In this approach, the fluctuating Gunn–Peterson approximation (FGPA) is applied to the dark matter (DM) density to estimate the neutral hydrogen absorption. Occasionally, to assess the importance of hydrodynamical processes, smaller hydrodynamical simulations with or without feedback processes have been used (the Illustris simulation with a box size of 75 $h^{-1}$cMpc in Horowitz et al. 2021b, and the Nyx simulation with a box size of 100 $h^{-1}$cMpc in Krolewski et al. 2018, Horowitz et al. 2019, and Li et al. 2021).

Simulations show that some protoclusters host halos as massive as $\sim 10^{13.5} \, h^{-1} M_\odot$; therefore, it is expected that energetic galactic feedback processes, particularly those associated with active galactic nuclei (AGNs), will to some extent heat, ionize, and displace the gas within protoclusters. In this work, we take a step forward by using the state-of-the-art large hydrodynamical simulation TNG300-1, which incorporates detailed subgrid models for feedback processes (Pillepich et al. 2018a). We then compare our results with the FGPA prescription by applying it to the dark-matter-only counterpart of TNG300-1, which is identical except for its lack of treatment of hydrodynamical processes.

The organization of this paper is as follows. We describe the simulations and mock observations in Section 2. In Section 3, we introduce our method for detecting protoclusters/groups within the mock-observed tomography maps. We discuss the masses of these structures in Section 4, including estimates of the structures' masses at $z = 2.5$ and the anticipated collapsed masses at $z = 0$. We conclude in Section 5 by summarizing and discussing the applications of our findings.

## 2. Simulations and Mock Observations

### 2.1. Simulations

We use the state-of-the-art hydrodynamical simulation TNG300-1 (Marinacci et al. 2018; Naiman et al. 2018; Nelson et al. 2018; Pillepich et al. 2018b; Springel et al. 2018), run with the AREPO code. The hydro solver is based on the moving mesh approach (Springel 2010), which evolves the gas particles on a moving mesh constructed from a Voronoi tessellation. The simulation box size is 205$h^{-1}$ cMpc, and it contains 2500³ dark matter particles and a similar number of gas cells, resulting in a dark matter particle mass of $5.9 \times 10^7 \, h^{-1} M_\odot$ and an average mass of $1.1 \times 10^7 \, h^{-1} M_\odot$ for the gas cells. The gravity is modeled using the TreePM algorithm with the gravitational force resolution set to 1 $h^{-1}$ckpc.

Supernovae and black hole (BH) feedback are constrained to reproduce observable properties of today's galactic population such as star formation histories, the stellar mass function, the stellar and gas versus halo mass relations, and the BH mass versus stellar mass relation. Due to the large spatial scales of interest in our work (larger than ≈1 physical Mpc, or pMpc), the most relevant type of feedback is the energetic feedback from BH accretion. BH feedback is implemented as thermal energy injection in the high-accretion mode and kinetic winds in the low-accretion mode (Weinberger et al. 2017). The simulations do not solve the radiative transfer equations; therefore, the local photoionization in the proximity zone of AGNs is not modeled. However, local heating due to radiative BH feedback is enforced by modifying the cooling rates (Section 2.6.4 in Vogelsberger et al. 2013). To constrain the importance of hydrodynamical processes and galactic feedback, we also apply the FGPA to a dark-matter-only companion run of TNG300-1, which has the same initial conditions and number of dark matter particles. In this paper, we use the particle snapshots, (sub)halo catalogs, and merger trees provided by the IllustrisTNG collaboration.

## 2.2. Simulated IGM Absorption

We create mock-observed spectra by modeling the absorption from gas particles in the full-hydro TNG300-1 simulation. The spectra are generated using the `fake_spectra`[4] package (Bird et al. 2015; Bird 2017). In this work, we added Message Passing Interface (MPI) support, allowing the analysis to scale to many nodes, and make it possible to post-process large cosmological simulations. Briefly, `fake_spectra` calculates the absorption spectra for any ion in the simulation, including the neutral hydrogen used in this work, along thousands of sightlines in the simulation box. Each particle is approximated as an individual absorber giving rise to absorption characterized by a Voigt profile. The internal physical quantities of these cells are smoothed by the kernel appropriate for the simulation. The supported interpolation kernels are SPH, top-hat, and partial Voronoi reconstruction. For the `Arepo` simulations, we used a top-hat kernel with the same volume as the Voronoi cell for each particle. This approximation has been tested to reproduce the spectra created by a full Voronoi mesh reconstruction for redshifts $z < 5$ (Wu et al. 2019). The neutral hydrogen fraction is obtained by assuming ionization equilibrium in the presence of a uniform ultraviolet background (UVB). We solve the collisional/photoionization and recombination rate network equations provided in Katz et al. (1996). The self-shielding of dense neutral hydrogen is taken into account by equation A.8 in Rahmati et al. (2013), which is a fitting formula to a radiative transfer model. The assumed uniform photoionization rate dictates the mean absorbed flux in the simulated spectra. We rescale the optical depth in the spectra to enforce the empirical mean flux evolution, corrected for metal absorption (there is no metal absorption in the simulated spectra), as derived by Faucher-Giguère et al. (2008):

$$\bar{F} = \exp(-1.330 \times 10^{-3} \times (1 + z)^{4.094}). \quad (1)$$

The simulated spectra extend along the third dimension of the snapshot with a pixel width of $\approx 6.4$ km s$^{-1}$ to capture the effect of saturation. Once the high-resolution spectrum is extracted, the absorption is averaged over every 20 adjacent pixels to mimic the detector's pixel size in the LATIS ($\approx 129$ km s$^{-1}$). Further, we mimic the spectral broadening in LATIS data by smoothing each spectrum with a Gaussian kernel of $\sigma = 2.06$ Å in the observed frame.

To constrain the importance of simulated baryonic processes, we also apply the FGPA to a DM-only companion simulation with identical initial conditions to TNG300-1. This DM-only companion evolves cold dark matter (CDM) particles and baryons as a single collisionless fluid; therefore, it does not capture galactic feedback effects. In this approach, by applying cloud-in-cell (CIC) interpolation, the dark matter density and line-of-sight peculiar velocity fields are deposited onto a uniform grid of length $\sqrt[3]{N_{DM}}$, where $N_{DM} = 2500^3$ is the number of dark matter particles in the simulation. We use the efficient NBODYKIT[5] code (Hand et al. 2018) for this step. To mimic the baryonic pressure smoothing and estimate the baryonic density with this approach, it is often advised to smooth the DM density and velocity fields (e.g., Sorini et al. 2016). However, we verified that for simulating the neutral hydrogen absorption on scales of 1 pMpc, this initial smoothing is unimportant, confirming the assumption in previous works (Stark et al. 2015b; Newman et al. 2020). The neutral hydrogen density is related to this approximate baryonic density by assuming ionization equilibrium with a uniform UVB. The recombination and photoheating rates are functions of temperature, which in turn is a function of density. We connect density to temperature using the slope of the power law measured by Rudie et al. (2012). After including the Hubble flow and peculiar velocity of the cells, we obtain the optical depth of the neutral hydrogen along the line of sight by convolving the density with a thermal Doppler broadening profile. Finally, we scale this optical depth to match the observed mean flux by Faucher-Giguère et al. (2008). The `Python` implementation is accessible through our repository.[6]

Moreover, in Appendix A we show both simulated full-hydro TNG300-1 and the FGPA method recover the 1D flux power spectra with a $\sim 10\%$ accuracy when compared to Sloan Digital Sky Survey (SDSS) DR14 results (Chabanier et al. 2019).

### 2.3. Realistic Mock Spectra

From these noiseless spectra, we produce a set of realistic mock spectra that match the areal density of sightlines in LATIS and their observational noise, which are the key parameters determining the quality of the derived tomographic maps (Stark et al. 2015b; Newman et al. 2020). As a representative value, we consider the average sightline density in the COSMOS field (now fully observed) in the redshift range $z = 2.4$–2.6 (the middle of the survey), which is $n = 0.16$ $\bar{h}^2$ cMpc$^{-2}$. This corresponds to a mean sightline separation of $\langle d_\perp \rangle = n^{-1/2} \approx 2.70$ $h^{-1}$cMpc and a total of 6739 mock spectra within the TNG300-1 simulation box. The variations due to redshift dependencies of these parameters within the observed range of LATIS are discussed in Appendix C.

We randomly select these sightlines and assign each a continuum-to-noise ratio (CNR) drawn from the distribution in LATIS (Newman et al. 2020). The CNR distribution is different for LBGs and QSOs, which constitute 98% and 2% of sightlines, respectively. In the mock observations, we randomly divide the sightlines into these two populations. The CNR is modeled as a log-normal distribution with a constant mean and variance for QSO background sources, while for LBGs, the variation of the mean with redshift is modeled with a second-degree polynomial. These equations are summarized in Table 1. The error propagated by continuum fitting is also implemented as prescribed by Krolewski et al. (2018):

$$F_{obs} = \frac{F_{sim}}{1 + \delta_{cont}}, \quad (2)$$

where $\delta_{cont}$, which is constant along each sightline, is a realization of a normal distribution $\mathcal{N}(0, \sigma_{cont})$ with $\sigma_{cont} = 0.24 \times$

**Table 1**
Continuum-to-noise Ratio Distribution of LBGs and QSOs, Parameterized as Log-normal Distributions

| Source | Mean | $\sigma$ |
|---|---|---|
| LBGs | $0.84 + 0.99$ $(z - 2.5) - 1.82(z - 2.5)^2$ | 0.43 |
| QSOs | 2.3 | 1.2 |

**Note.** The mean and standard deviation of ln(CNR) are listed.

---

[4] https://github.com/sbird/fake_spectra
[5] https://nbodykit.readthedocs.io
[6] 10.5281/zenodo.6130050

CNR$^{0.86}$ (Newman et al. 2020). Finally, we follow the prescription of Newman et al. (2020) and remove lines with an equivalent width (EW) > 5 Å, which is intended to suppress damped absorption lines. We emphasize that rather than masking such lines explicitly based on the total neutral hydrogen column density along the sightline, we closely mimic the LATIS procedure by applying it to the mock spectra.

### 2.4. Generating 3D Flux Maps

To generate the 3D Ly$\alpha$ mock absorption map, we first create a random set of mock-observed spectra at $z = 2.45$, close to the mid-redshift of the LATIS survey. In Appendix C, we discuss our results for slightly different redshifts within the observed range in LATIS. However, the remainder of the paper is based on the snapshot at $z = 2.45$ (hereafter abbreviated $z = 2.5$). The quantity of interest is the flux contrast $\delta_F$, defined as:

$$\delta_F = \frac{F}{\bar{F}} - 1, \tag{3}$$

where $\bar{F}$ is the mean flux. The flux contrasts in the mock spectra are interpolated to make a 3D map of $\delta_F$ using a Wiener filter. The Wiener filter is an optimal linear estimator of the 3D $\delta_F$ map from a set of observed noisy spectra and specified noise and signal covariance matrices. It requires the noise to be additive; hence, the data vector is $\boldsymbol{n} = \boldsymbol{s} + \boldsymbol{n}$, where $\boldsymbol{s}$ and $\boldsymbol{n}$ are the vectors of signal and noise, respectively. By minimizing the mean squared error between the linear estimate, $\hat{s} = \boldsymbol{L}\boldsymbol{d}$, and the true signal, one obtains $\boldsymbol{L} = \boldsymbol{C}_{mp}(\boldsymbol{C}_{pp} + \boldsymbol{N})^{-1}$. $\boldsymbol{C}_{mp}$ and $\boldsymbol{C}_{pp}$ are the assumed signal covariance matrices, where the $m$ and $p$ indices indicate voxels in the 3D map and pixels along the sightlines, respectively. $\boldsymbol{N}$ is the noise covariance, which is assumed to be diagonal. Following the *ad hoc* assumption in Caucci et al. (2008; also in Stark et al. 2015b), we set $\boldsymbol{C}_{pp} = \boldsymbol{C}_{mp} = \boldsymbol{C}(\boldsymbol{r}_1, \boldsymbol{r}_2)$ as a Gaussian:

$$\boldsymbol{C}(\boldsymbol{r}_1, \boldsymbol{r}_2) = \sigma_F^2 \exp\left[-\frac{\Delta r_\parallel^2}{2\sigma_\parallel^2} - \frac{\Delta r_\perp^2}{2\sigma_\perp^2}\right], \tag{4}$$

where $\Delta r_\parallel$ and $\Delta r_\perp$ are the components of $\boldsymbol{r}_1 - \boldsymbol{r}_2$ parallel and perpendicular, respectively, to the line of sight. The amplitude is set to $\sigma_F^2 = 0.05$, and the correlation lengths are fixed at $\sigma_\perp = 2.5$ $h^{-1}$cMpc and $\sigma_\parallel^2 = \sigma_\perp^2 - \sigma_{inst}^2$, where $\sigma_{inst} = 1.4$ $h^{-1}$cMpc is the instrumental resolution of LATIS at $z = 2.5$. The diagonal entries of the noise covariance matrix, $\boldsymbol{N}$, are sourced by contributions from both random noise and continuum fitting error, $\sigma_\delta = (\sigma_{rand}^2 + \sigma_{cont}^2)^{1/2}$, including a floor of $\sigma_\delta > 0.2$ on the total noise. All of these parameters match those applied to the LATIS observations by Newman et al. (2020). We use DACHSHUND,[7] an efficient Wiener filtering implementation by Stark et al. (2015b).

Figure 1 visually compares the quality of the flux map recovered from the realistic mock observations to that of a noiseless mock map and the underlying dark matter density field. The noiseless mock map (middle column) is produced by generating spectra on a regular grid with a fine spacing of $1$ $h^{-1}$cMpc and adding no noise; the noiseless map is not

Wiener filtered. All maps are smoothed with a Gaussian kernel with length of $\sigma = 4$ $h^{-1}$ cMpc, the typical scales of protoclusters at $z = 2.5$ (Chiang et al. 2013; Stark et al. 2015b). We follow the convention in previous works of normalizing $\delta_F$ by its standard deviation $\sigma_{map}$ over the map volume; note that $\sigma_{map}$ is not the measurement error. The visual similarity among these maps and the voxel-by-voxel comparison in the bottom panel of Figure 2 confirm the quality of the flux map recovered on megaparsec scales.

The optimal reconstruction of the 3D absorption field from a discrete set of noisy spectra is an active field of research. Pichon et al. (2001) compared a constrained Gaussian random field linear approach (or Wiener filter) with a general nonlinear explicit Bayesian deconvolution method. They show that for low-resolution spectra with negligible measurement errors, and under the ad hoc assumption of Gaussian distributed flux, Wiener filtering produces the maximum a posteriori map of the Bayesian method. Using new optimization tools, Li et al. (2021) loosened the constraint that the flux correlation is Gaussian and minimized a loss function for a more general signal. In parallel, some other interesting methods under development aim at simultaneous reconstruction of matter density and velocity fields via reconstructing the likeliest initial conditions (Horowitz et al. 2019, 2021b; Porqueres et al. 2020). However, in this work, we adopt the Wiener filter approach for reconstructing the Ly$\alpha$ flux field, since it is used by current-generation surveys, including LATIS. We defer evaluating other methods to future work.

#### 2.4.1. Flux pdf

In order to verify that the $\delta_F$ fluctuations in the mock-observed maps are consistent with those seen in LATIS, we compared the probability distribution functions (pdfs). For this purpose, we use the LATIS map in the COSMOS field, updated from the version presented in Newman et al. (2020) to include the fully observed volume. To account for cosmic variance, the mock map is cut into a few subboxes, each with roughly the same volume as LATIS. Figure 3 shows the flux pdf in LATIS ($z = 2.2$–2.8, dashed black) and mock-observed maps ($z = 2.45$, orange stripes enclosing 68% and 95% of the subboxes) for three different smoothing scales. The agreement is excellent. We verified that this agreement holds when the LATIS pdf is computed in several narrower redshift bins ($z = 2.2$–2.4, 2.4–2.5, 2.5–2.8) and compared to the nearest TNG snapshot. Figure 3 demonstrates the overall agreement between the mock-observed signal predicted by $\Lambda$CDM and observed in LATIS.

#### 2.4.2. Correlation with the Dark Matter Density

Quantifying the correlation between the mock-observed signal $\delta_F$ and the underlying density field on a voxel-by-voxel basis is an essential first step toward estimating the masses of structures (Lee et al. 2016). Since the flux map, which will be used to define the boundaries of structures, is reconstructed in velocity space, we computed the density of dark matter particles in velocity space. Specifically, the particles were displaced by their line-of-sight peculiar velocities and then interpolated with a CIC kernel onto a regular mesh grid with a resolution of $1$ $h^{-1}$cMpc. The top panel in Figure 2 illustrates the correlation between these two fields, $\delta_F^{sm}$ and $\rho_{DM}^{sm}$, both smoothed by a $\sigma_{sm} = 4$ $h^{-1}$cMpc Gaussian at

---

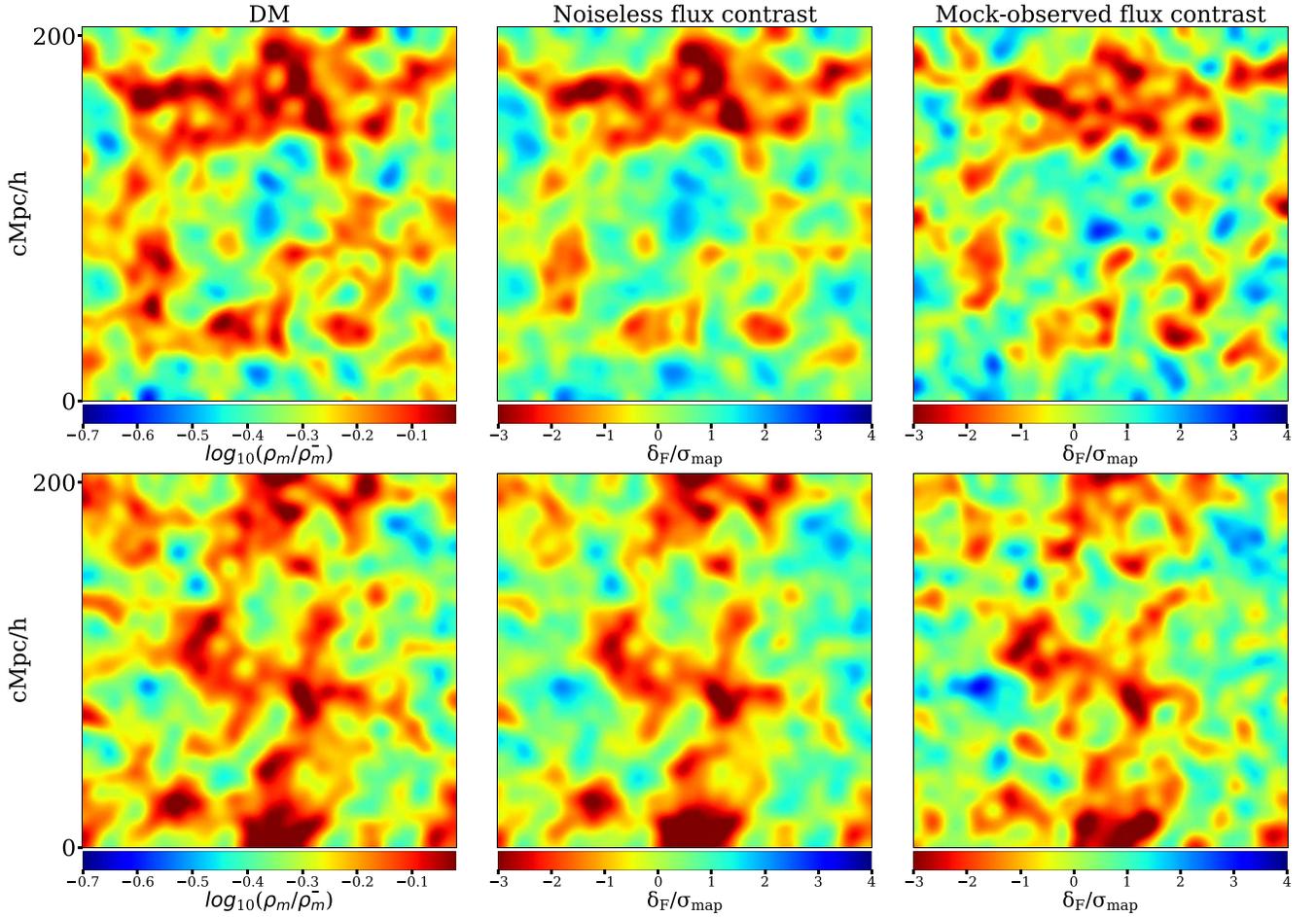

**Figure 1.** Visualizations of two different slices (top and bottom rows) of the IllustrisTNG-300 hydro simulation at $z = 2.45$. All maps are smoothed by a Gaussian with kernel size $\sigma_{sm} = 4\ h^{-1}$ cMpc. Left column: the smoothed DM density, $\rho_{DM}/\overline{\rho_{DM}}$. Middle column: the smoothed noiseless Ly$\alpha$ flux contrast $\delta_F/\sigma_{map}$. Right column: the mock-observed flux contrast, constructed by Wiener interpolation of spectra that mimic LATIS observations. The mock $\delta_F$ maps trace the underlying DM distribution very well on large scales.

$z = 2.5$. The solid curve is the median second-degree polynomial fit among 20 mock-observed maps. This estimator, along with its uncertainties, is

$$\left(\frac{\rho_{DM}}{\langle \rho_{DM} \rangle}\right)^{sm}(z = 2.5) = 16.78(\pm 0.53)\ \delta_F^{sm\ 2}$$
$$- 5.62(\pm 0.07)\ \delta_F^{sm}$$
$$+ 0.96(\pm 0.01). \quad (5)$$

In Appendix C we show that although this relation evolves slightly within the redshift range of LATIS, its evolution is negligible for the purposes of this paper.

## 3. Characterizing Structures in Ly$\alpha$ Tomographic Maps

Ly$\alpha$ tomography provides continuous maps of the high-redshift H I absorption field and, via the correlation shown in the top panel of Figure 2, of the matter density field. However, it is often useful to identify and characterize discrete structures in these maps. Dark matter halos at $z \approx 2.5$ do not provide a functional route to define these structures, since due to hierarchical structure formation, the resolution of the maps is much coarser than the size of even the most massive halos yet formed. Instead, most previous works have focused on the volumes that will collapse into massive halos at $z = 0$, which are generally called protoclusters when the descendant mass $M_{z=0} > 10^{14}\ h^{-1} M_\odot$. This perspective has the merit of supplying a rigorous way to define these high-redshift "structures," but the connection between these structures and the observed $z = 2.5$ Ly$\alpha$ field is complex due to the structures' varied assembly histories. In other words, variations in the kinematics and density structure of subcomponents within detected overdensities affect their evolution such that not all absorption-detected structures will form a galaxy cluster by $z = 0$, and not all galaxy clusters will have a progenitor detectable in absorption at $z = 2.5$. Furthermore, estimating density and environment at high redshifts (not just the structure's future evolution) is important for understanding the impact of environment on galactic evolution.

In this section, we describe a method for identifying structures directly in Ly$\alpha$ tomographic maps. The method extends previous techniques (Lee et al. 2016) in two ways. First, we include a larger selection of structures that extends to lower masses. Second, we introduce a method to delineate boundaries between overlapping structures. These improvements will aid future galaxy evolution studies by providing both larger numbers of structures and a simple way to associate galaxies with them. In Section 4, we will consider the masses of these structures and their descendants.

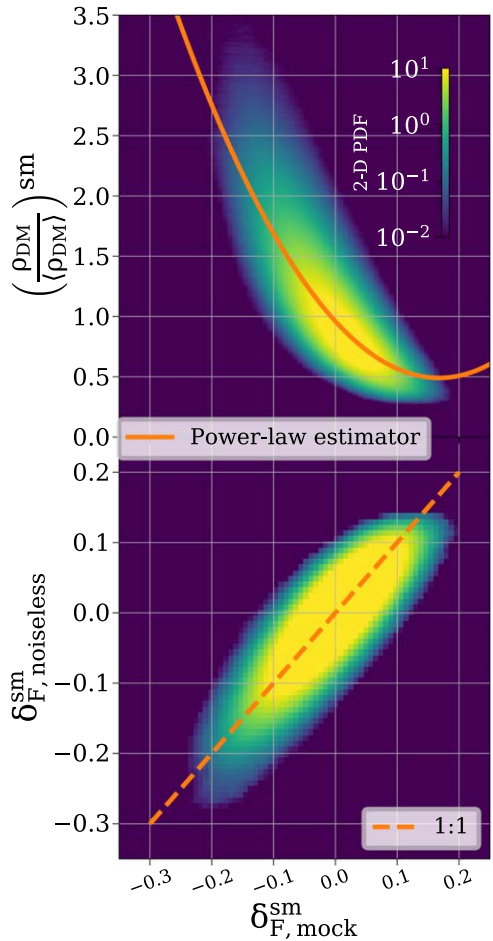

**Figure 2.** Upper panel: the median correlation between the smoothed ($\sigma_{sm} = 4\ h^{-1}$cMpc) DM density and the mock-observed flux contrast $\delta_{F,\,mock}^{sm}$ on a voxel-by-voxel basis in 20 mock maps at $z = 2.5$. The solid line is the median second-degree polynomial estimator (Equation (5)), which we use to estimate the DM mass within structures detected in the tomographic maps. Lower panel: a voxel-by-voxel comparison between the noiseless and mock-observed flux contrasts, smoothed as in the upper panel. In both panels, median 2D histograms with logarithmic color scales are displayed.

### 3.1. Defining Detected Structures as Flux Islands

Our basic unit for a structure in a smoothed tomographic map is a contiguous region in which the flux is below a threshold $\delta_F/\sigma_{map} < \nu$, which we call an "island." By default, we consider maps smoothed with a Gaussian $\sigma_{sm} = 4\ h^{-1}$cMpc since the map signal-to-noise ratio degrades significantly when less smoothing is applied (see Figure 22 in Newman et al. 2020). Stark et al. (2015b) showed that this smoothing scale is nearly optimal for the detection of protoclusters. In Appendix D, we show that some lower-mass structures leave significant signatures only in less-smoothed maps, but due to larger uncertainties in characterizing those structures, we will focus only on the $\sigma_{sm} = 4\ h^{-1}$cMpc smoothed maps for the rest of this paper.

To determine the optimal value of $\nu$, we need to balance completeness, which favors a high $\nu$ (less negative and less absorption) to include more structures, with purity, which favors a low $\nu$ (more significant and more absorption) to suppress the influence of noise. To estimate the level of noise in the maps, we generated a pure noise map by mock observing a structureless field (i.e., adding noise to $\delta_F = 0$ to create the spectra, then Wiener filtering as described in Section 2.4). Figure 3 shows the flux pdf of the voxels in the mock-observed and the pure noise maps. The ratio of the two distributions at $\nu \approx -2$ is $\approx 0.01$, indicating that this threshold selects voxels where absorption ($\delta_F < 0$) is present at >99% confidence. Furthermore, we will show in the next section that, after appropriate subdivision, these islands are directly connected to the descendant massive halos at $z = 0$ only for a small range of $\nu$, and any deviation from this leads to either the emergence of unrealistically large structures or missing some significant structures in the map.

### 3.2. Deblending Substructures in Flux Islands with the Watershed Algorithm

Some flux islands extend over very large volumes, which may contain substructures (i.e., multiple flux minima) that will often collapse into several distinct halos. In these situations, it is useful to consider the island as being composed of multiple structures. Here we present a method to partition flux islands into smaller blocks, which we call "watersheds," each of which is associated with a distinct flux minimum. We will then analyze the connection between these watersheds and the progenitors of massive halos at $z = 0$.

Figure 4 shows a slice through our mock map at the location of a large contiguous region with $\delta_F/\sigma_{map} \leqslant -2$ that contains several local minima. Diamonds show the centers of massive protogroups and protoclusters. These are computed by tracing back the DM particles located within $R_{200}$ of a massive halo ($M_{vir} > 10^{13.5}\ h^{-1}M_\odot$) at $z = 0$; the center is then identified as the center of mass of these particles. As expected, the progenitors of massive halos cluster strongly; this flux island contains the progenitors of several massive halos. The flux minima do not always have a one-to-one correspondence with massive halo progenitors, both because of observational noise and the fact that the progenitors are blended at the map resolution. Thus, we do not expect that it will be possible to cleanly deblend all massive halo progenitors in our maps. Nonetheless, Figure 4 makes it clear that dividing islands into watersheds associated with local flux minima could provide a simple and effective way to deblend many of the massive protogroups and protoclusters.

To divide the flux islands into smaller blocks, we use the watershed algorithm (van der Walt et al. 2014), hence naming the blocks "watersheds." This algorithm has been used previously in other cosmological applications (e.g., for detecting voids; Platen et al. 2007). The watershed algorithm floods basins from a set of markers until basins associated with different markers meet on the watershed line. We flood flux islands and take a subset of the local flux minima as the markers. The subset is defined by a threshold on the flux contrast $\delta_F/\sigma_{map} \leqslant \kappa$. Both $\kappa$ and $\nu$ are selectable parameters in our method.

In Section 3.1, we motivated $\nu \approx -2$ based on the noise level in the mock-observed maps. An additional desirable property of the watersheds is that they approximately enclose the Lagrangian extent of the progenitors of $z = 0$ massive halos. To investigate this, we varied $\nu$ and subdivided islands into watersheds, considering all of the local minima within each island, i.e., setting $\kappa = \nu$. We then integrated the estimated dark matter density $\rho_{DM}$, as estimated from the flux contrast $\delta_F$ via Equation (5), over each watershed's volume $W$ to form a raw

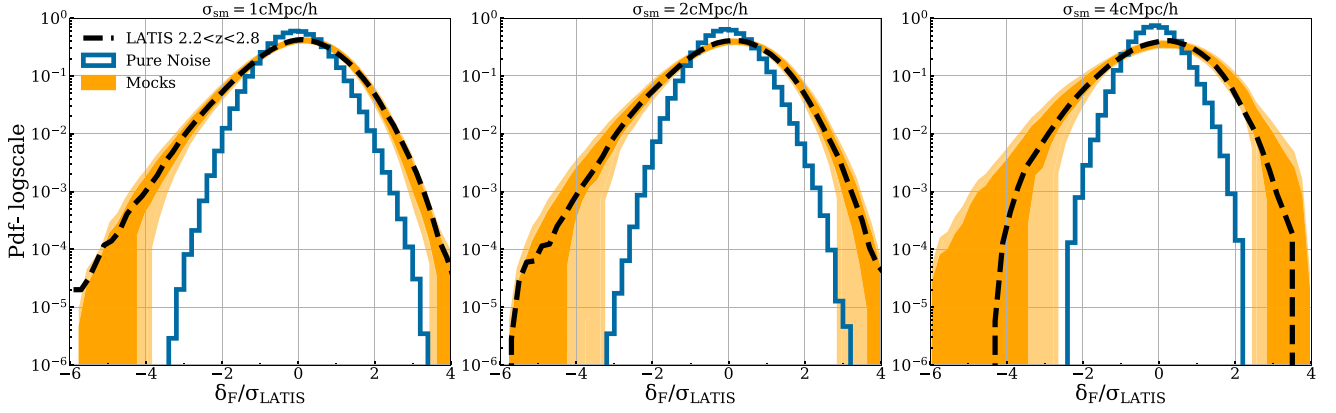

**Figure 3.** The normalized distribution of the flux contrast, smoothed on three different scales, in LATIS and the mock maps. We normalize $\delta_F$ by its standard deviation $\sigma_{LATIS}$ in the LATIS map. The black dashed lines show the pdfs of the observed map, and the orange stripes show those of mock-observed maps at $z = 2.45$. The width of the stripes represents the cosmic variance within 68% and 95% confidence levels. The solid blue curves show the pdfs of pure noise maps generated by Wiener filtering a set of pure noise spectra (see Section 3.1). These plots confirm the accuracy of the mock maps in reproducing the basic statistics of the LATIS maps.

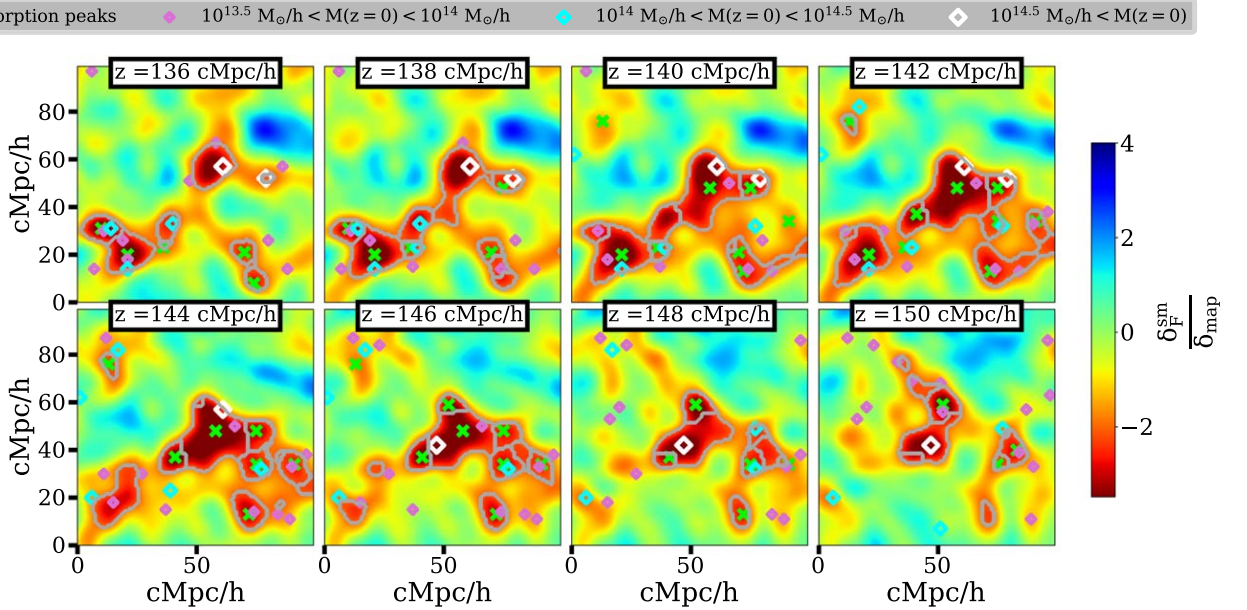

**Figure 4.** Illustration of the watersheds identified in a portion of the mock-observed map and their connection to the progenitors of massive halos at $z = 0$. The $x$, $y$, and $z$ coordinates are those of the simulation box. Crosses mark the absorption peaks in the mock-observed map, and diamonds represent the centers of mass of the progenitors of massive halos at $z = 0$. For better visibility, these markers are copied along $\Delta z = \pm 4 \ h^{-1}$cMpc of the actual location. The size and color of diamonds indicate the $z = 0$ halo mass. Silver contours illustrate the watersheds constructed using our method (Section 3.2), which better isolates progenitors in highly blended regions by using the gradient information in the flux map.

tomographic mass:

$$M_{\rm tomo, \, raw} = \int_W \left( \frac{\rho_{\rm DM}}{\langle \rho_{\rm DM} \rangle} \right)^{\rm sm} (\delta_F) \times \langle \rho_{\rm DM} \rangle (z) \, dV. \quad (6)$$

We refer to this as a "raw" tomographic mass to distinguish it from the calibrated mass $M_{\rm tomo}$ that we will define in Section 4. Figure 5 compares the distribution of $M_{\rm tomo, \, raw}$ to the halo mass function at $z = 0$. For $\nu > -2$, the watersheds become very large and contain much more mass than equally numerous halos at $z = 0$. On the other hand, setting $\nu < -2$ produces islands that do not contain all of the mass that will collapse. In addition to underestimating the volumes of protogroups and protoclusters, choosing $\nu < -2$ will also exclude some structures that produce significant absorption in the maps.

Therefore, we define flux islands as regions with $\delta_F / \sigma_{\rm map} \leqslant \nu = -2$.

In order for the volume and mass of an island to be well defined, the absorption must reach a minimum depth below the contour level $\nu$ used to define the island. Otherwise, the volume can diverge to 0, and the uncertainties in the mass estimates become too large to be useful. We impose this margin by requiring flux islands to have a local minimum with $\delta_F / \sigma_{\rm map} \leqslant \kappa$. When an island has multiple such minima, we use only the minima with $\delta_F / \sigma_{\rm map} \leqslant \kappa$ as "markers" in the watershed algorithm. Thus, each watershed is associated with exactly one such local minimum.

In order to set the value of $\kappa$, we minimize the error in the tomographic masses of the watersheds. We compute $M_{\rm tomo,raw}$ over the watersheds obtained using $\nu = -2$ and several values for $\kappa$. The error is then defined as the difference between counterpart watersheds in the mock-observed and the noiseless flux maps. The counterpart of a mock-observed watershed is the one in the noiseless map with the largest overlap.[8] Finally, for a given value of $\kappa$, we calculate the rms of these errors over all watersheds. Figure 6 shows the rms as a function of $\kappa$. Choosing $\kappa < \nu = -2$ reduces the rms mass error considerably. But as $\kappa$ is progressively lowered, the improvement in the mass error diminishes, and more structures are excluded from the analysis. Balancing these considerations, we select $\kappa = -2.35$. Coincidentally, this value matches the threshold used by Newman et al. (2020) to select voxels with a 95% probability of lying in the bottom 10% of the $\delta_F$ distribution.

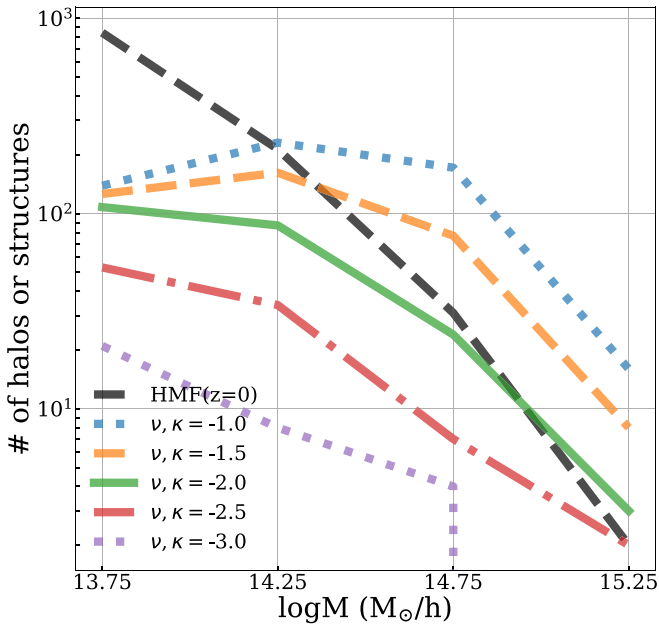

**Figure 5.** The distribution of the watersheds' tomographic masses $M_{tomo,raw}$, for several choices of the contour level $\nu$, is compared to these contour at $z = 0$. For this comparison, we define watersheds using all of the absorption peaks and hence set $\kappa = \nu$. The tomographic masses are well matched to the masses of protoclusters when $\nu = -2$, which motivates our choice of this parameter.

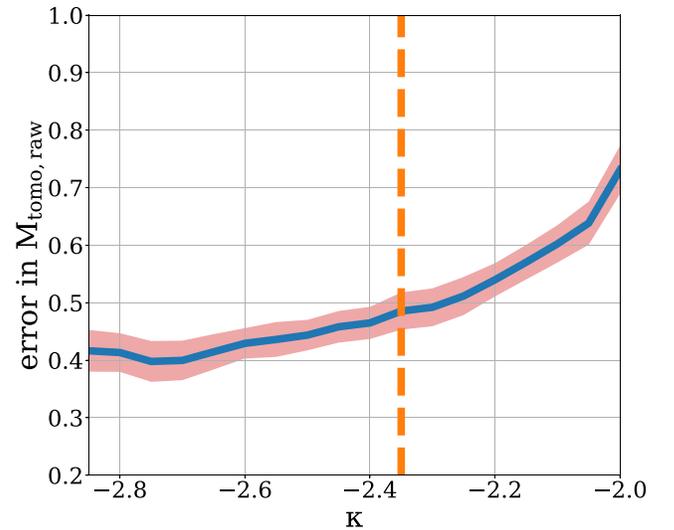

**Figure 6.** The rms error in the tomographic masses $M_{tomo,raw}$ of watersheds as a function of the parameter $\kappa$. The error is defined based on the difference between masses of watersheds in the mock-observed map and their counterparts (based on maximal overlap) in the noiseless map. The solid line and shaded regions are the mean and $1\sigma$ scatter among different realizations of the mock-observed map. Based on this analysis, we select $\kappa = -2.35$.

A visual example of the watersheds produced by this method is shown in Figure 4. The centers of mass of the progenitors of massive $z = 0$ halos are indicated with diamonds, with size and color depending on the descendant halo mass. This visually confirms the capability of the method to deblend massive halo progenitors located within the same island. Figure 7 quantifies this by comparing the number of progenitors within each island or watershed. The distribution of progenitors for islands, contours of $\delta_F/\sigma_{map} \leqslant \nu = -2$, shows that many of them host no massive halo progenitor, at least in the mass range we consider here, $M_{z=0} > 10^{13.5} \, h^{-1} M_\odot$. Many of these islands are removed after requiring the islands to have at least one minimum with $\delta_F/\sigma_{map} \leqslant \kappa = -2.35$. However, there are some very large islands hosting up to 40 progenitors (the last bin of the histogram). By using the information on the gradient of the flux field, our method breaks these giant islands into watersheds that isolate progenitors individually, or at least in much smaller groups. This comparison shows that watersheds are clearly superior at isolating the progenitors of massive $z = 0$ halos compared to simple contours in the flux map. Nevertheless, some progenitors are extremely blended and cannot be distinguished even with our deblending method.

## 4. Results

In this section, we define and calibrate the tomographic masses $M_{tomo}$ of watersheds, which can be estimated from a tomographic map. We then compare these tomographic masses to the underlying dark matter mass at $z = 2.5$, and to the masses of halos at $z = 0$ that the watersheds will form. Our aim is to develop and characterize estimators for these masses that can be applied to tomographic surveys, including LATIS.

Before discussing these estimators, we first review our notation for the various masses that will be discussed. $M_{tomo}$ denotes a mass calculated from a flux map. This could be a mock-observed map or a noiseless one, but for the rest of the paper, unless otherwise specified, we will consider $M_{tomo}$ measured in mock-observed maps. When we consider the dark matter mass of a watershed at $z = 2.5$, we find that it is important to distinguish the mass within the watershed boundary derived from the mock-observed map, which we denote $M_{DM, mock}$, from the mass within the boundary derived from a noiseless flux map, which we denote $M_{DM, noiseless}$. $M_{desc}$ will refer to the virial mass at $z = 0$ of the halo identified as the main descendant of a watershed. Finally, we use "est" to distinguish estimators, which are functions of $M_{tomo}$. Thus $M_{DM}^{est}$ and $M_{desc}^{est}$ are estimators of $M_{DM,noiseless}$ and $M_{desc}$, respectively.

### 4.1. Calibrating Tomographic Masses at $z = 2.5$

As introduced in Section 3.2, $M_{tomo,raw}$ is defined by estimating the dark matter density at each voxel, using the median $\delta_F^{sm}$–$\rho_{DM}^{sm}$ relation (Equation (5)), and integrating over the watershed volume. As discussed by Lee et al. (2016), using a smoothed density field pushes some particles out of the watershed volume, which leads to a systematic underestimation of the DM mass within watersheds. We correct for this by applying an offset to our tomographic mass estimates, which are defined as

$$M_{\rm tomo} = 10^{0.14} \times M_{\rm tomo,raw}. \tag{7}$$

---

[8] The amplitude of the fluctuations in the mock-observed and noiseless flux maps are slightly different. In the mock-observed map, it is regulated by the signal covariance function supplied to the Wiener filter, $\sigma_F$ in Equation (4); therefore, it is necessary to first scale the noiseless map to match the rms of the mock-observed one.

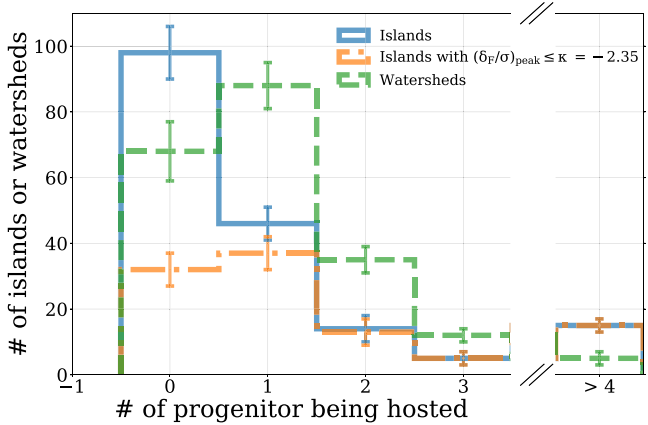

**Figure 7.** Comparing the utility of watersheds and simple flux contours (i.e., islands) in detecting the progenitors of massive $z = 0$ halos. Histograms show the mean distributions of the progenitor counts within islands or watersheds in 20 mock-observed flux maps. The $1\sigma$ scatter around mean counts are also overlaid. Many islands, contours of $\delta_F/\sigma_{\rm map} \leqslant \nu = -2$ (blue solid curve), host no progenitors exceeding $M_{z=0} > 10^{13.5}\ h^{-1}M_\odot$. Requiring these islands to contain at least one absorption peak with ($\delta_F/\sigma_{\rm map} \leqslant \kappa = -2.35$) removes many of these vacant islands (orange dotted–dashed curve), but still leaves islands that contain as many as 40 progenitors. The watershed algorithm partitions those islands, significantly increasing the number of structures containing two or fewer progenitors (green dashed curve).

The motivation for the calibration factor $10^{0.14}$ is as follows. In the left panel of Figure 8, we compare $M_{\rm tomo}$ to the dark matter mass $M_{\rm DM,\ mock}$ (i.e., the integral of the unsmoothed DM density) within the same watershed contours.[9] The slope of the linear fit is very close to one, and with our choice of calibration factor in Equation (7), there is no offset. The very small scatter of 0.12 dex originates from the scatter in the $\delta_F^{\rm sm}$–$\rho_{\rm DM}^{\rm sm}$ relation. This comparison shows that $M_{\rm tomo}$ provides a remarkably precise estimate of the dark matter mass within a specified volume.

Due to noise in the mock observations, which originates from the finite signal-to-noise and sightline density, the boundaries assigned to watersheds are distorted relative to those in the noiseless map. The right panel of Figure 8 compares the tomographic masses $M_{\rm tomo}$ of watersheds in the mock-observed map to the dark matter mass of companion watersheds in the noiseless map, $M_{\rm DM,\ noiseless}$. Again, the companion watershed is defined as the one that maximally overlaps a mock-observed one. Around 18% of watersheds in the mock-observed map overlap with no structure in the noiseless map; hence, they are not shown in this figure. This comparison incorporates the additional uncertainty arising from the noisy boundaries of watersheds in the mock-observed map, which greatly inflates the scatter and produces a mass-dependent offset in the $M_{\rm tomo}$–$M_{\rm DM,noiseless}$ relation. In order to define an estimate for the expectation value of $M_{\rm DM,noiseless}$ given an observed $M_{\rm tomo}$, we fit a power law using ordinary least-squares regression in logarithmic space. We generate 20 mock-observed maps with different sightlines and noise realizations. We use 16 of these maps as the training set to estimate the mean and uncertainties in the fit parameters; the mean relation, shown as the dotted line in the right panel of Figure 8, is

$$M_{\rm DM}^{\rm est} = 10^{14.60 \pm 0.05} \times \left( \frac{M_{\rm tomo}}{10^{14}} \right)^{0.39 \pm 0.07} \quad h^{-1}M_\odot. \tag{8}$$

We use the remaining four mock-observed maps to measure the scatter of the dark matter mass $M_{\rm DM,\ noiseless}$ around the estimator $M_{\rm DM}^{\rm est}$. The rms deviations within mass bins of $M_{\rm tomo}/\ h^{-1}M_\odot = [10^{13}, 10^{14}], [10^{14}, 10^{14.5}], [10^{14.5}, 10^{15.5}]$ are 0.36, 0.33, and 0.29 dex. Therefore, the uncertainty in the mass estimate for watersheds is mostly driven by the noisy boundary of the watersheds rather than scatter in the $\delta_F^{\rm sm}$–$\rho_{\rm DM}^{\rm sm}$ relation.

### 4.1.1. Collisionless versus Hydrodynamical Simulations

Collisionless simulations are often used to estimate the flux field, because only they have enough volume to enable many independent realizations of current-generation tomographic surveys. However, these simulations exclude physics that may be important for accurately estimating the Ly$\alpha$ flux field on megaparsec scales. To assess the accuracy of this method, we compare the noiseless $\delta_F$ maps obtained from FGPA and the full-hydro simulation. The contours in Figure 9 show the global distribution of the smoothed ($\sigma_{\rm sm} = 4\ h^{-1}$cMpc) transmitted flux contrast versus the DM density, a diagnostic plot suggested in Kooistra et al. (2022a, 2022b) to measure the effect of feedback on Ly$\alpha$ tomographic maps. Contrary to Kooistra et al. (2022b), Figure 9 shows a very small difference between these two methods on scales relevant to our analyses. In Appendix B, we show how differences between our FGPA and hydro maps are affected by the smoothing scale. We demonstrate that differences are slightly larger with a 3 Mpc $h^{-1}$ smoothing scale, but still negligible compared to the current observational uncertainties in the $\delta_F^{\rm sm}$–$\rho_{\rm DM}^{\rm sm}$ relation.

Moreover, Figure 10 compares the tomographic masses of the watersheds detected in the FGPA and full-hydro noiseless maps. Counterpart watersheds are those with maximal overlap. The median difference in $M_{\rm tomo}$ within bins of $\log(M_{\rm tomo}/\ h^{-1}M_\odot) = [13.5, 14, 14.5, 15.5]$ is consistent with zero, confirming that the FGPA-based estimates are unbiased with respect to the full-hydro calculation. The scatter between the $M_{\rm tomo}$ estimates is generally small: $\sigma = 0.19, 0.18$, and 0.14 dex in the same mass bins, where $\sigma$ is the standard deviation obtained with an iterative sigma-clipped method. However, there is a noticeable set of outliers. These appear mainly in the left panel of Figure 10, which compares mass estimates for the watersheds detected in the full-hydro map, and are less numerous in the right panel, which shows the watersheds detected in the FGPA map. This comparison indicates that the main origin of the outliers is low-mass watersheds in the full-hydro map that are not present as separate structures in the FGPA map, because they are blended with larger watersheds (left panel). We find that 32 watersheds (18% of the total) in the full-hydro map are over-blended in the FGPA map (i.e., they share a counterpart), while a smaller number (16, which is 9%) are over-fragmented.

---

[9] We do not include the baryonic mass in $M_{\rm DM}$, but users who wish to estimate a total mass could simply scale by $\Omega_m/\Omega_{\rm DM}$, which is sufficiently accurate given the large scales we are considering.

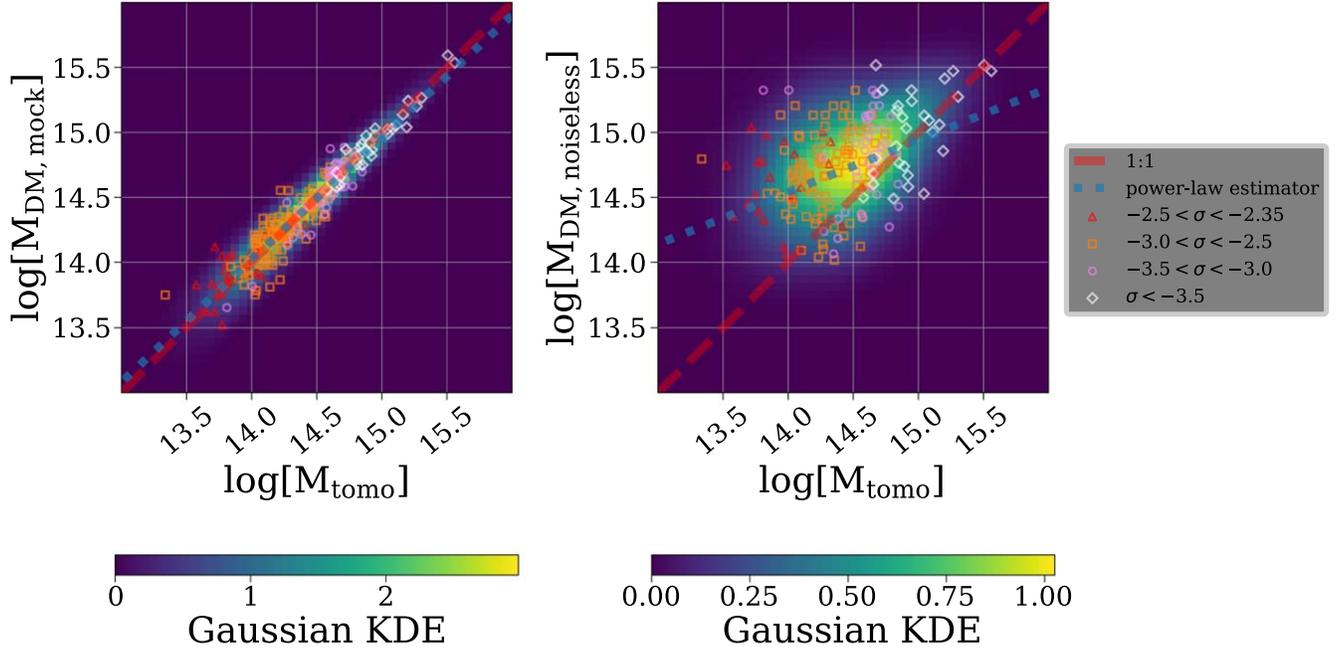

**Figure 8.** Left panel: the tomographic masses (Equation (7)) of watersheds in a mock-observed map are compared to the dark matter mass (i.e., the integrated unsmoothed dark matter density) within the same contours. The points are scattered around the one-to-one relation with a standard deviation of 0.12 dex. This scatter is sourced by scatter in the $\delta_F^{sm}$–$\rho_{DM}^{sm}$ relation. Right panel: the tomographic mass is compared to the dark matter mass within companion structures in the noiseless flux map. The dotted line is the mean power-law fit $M_{DM}^{est}$ obtained from 16 mock-observed maps (Equation (8)). The increased scatter and bias arise from noise in the boundaries of the watersheds. Points representing individual watersheds in one of the mock maps are overlaid on the mean Gaussian kernel density estimation (KDE) among 20 mock maps.

adequate for simulating the signal from more massive structures, around $M_{tomo} \gtrsim 10^{14.2} M_\odot$. However, it becomes concerning in some lower-mass structures; in particular, those located near massive structures are prone to over-blending.

### 4.2. The Connection to Massive Halos at $z = 0$

To enable statistical studies of galactic evolution over cosmic time, it is useful to relate the structures detected in tomography surveys to virialized massive halos in the late-time universe. Because the connection between the two is not one-to-one, tracing the descendants of high-$z$ structures is not equivalent to tracing the progenitors of $z = 0$ halos, and the appropriate direction depends on the question. Here we consider both approaches in turn.

#### 4.2.1. Descendants

In this approach, we associate every watershed in a mock-observed tomographic map with a halo at $z = 0$. We aim to find the halo at $z = 0$ where the majority of the mass within the watershed structure will end up. To do so, we first link all halos (friends-of-fiends groups) at $z = 2.5$ to a halo at $z = 0$. Specifically, halo **B** at $z = 0$ is the descendant of halo **A** at $z = 2.5$ if the descendant of the most massive subhalo within halo **A** lies within the halo **B**. This approach allows us to use the SubLink trees provided with IllustrisTNG to connect subhalos across different redshifts. Then, to find the halo at $z = 0$ in which the majority of the collapsed mass within each tomographically mapped structure will end up, all halos within that structure vote for their descendant halo at $z = 0$, weighted by their halo mass. The mass of the descendant halo with the highest vote is chosen as the descendant of that watershed.

The main panel in Figure 11 compares the masses of these descendant halos, $M_{desc}$, to the mock-observed tomographic

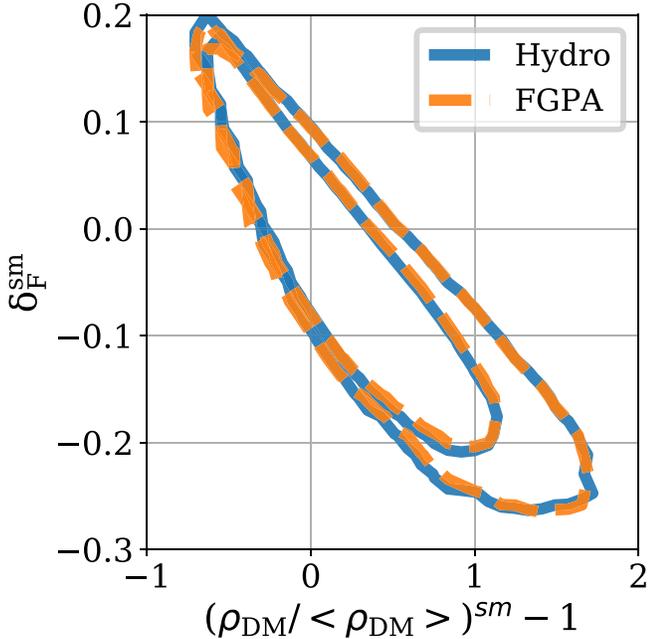

**Figure 9.** A comparison of the Ly$\alpha$ flux fields simulated using the FGPA and the TNG300-1 simulation. The FGPA is applied to a dark-matter-only simulation, while TNG300-1 includes hydrodynamics and state-of-the-art feedback models. The noiseless flux contrast and DM overdensity maps have a voxel resolution of 1 $h^{-1}$cMpc and are smoothed with a Gaussian kernel of size $\sigma_{sm} = 4\ h^{-1}$cMpc. The contours are the 0.02, 0.68, and 0.98 levels for the 2D PDF. The strikingly similar $\delta_F^{sm}$–$\rho_{DM}^{sm}$ relation confirms the insensitivity of the Ly$\alpha$ tomography signal, on megaparsec scales, to the feedback and detailed hydro prescription.

Comparing this scatter to the uncertainty in the tomographic mass estimation discussed earlier (Figure 8) and considering the fraction of failures, we find that the FGPA approach is masses, $M_{tomo}$, of all individual watersheds detected in the mock-observed map. The power-law fit in that plot provides an estimate of the descendant mass of a watershed that can be applied to real observations. Again, we use 16 mock-observed maps to estimate the mean relation and the uncertainties in the parameters:

$$M_{desc}^{est} = 10^{13.72 \pm 0.09} \left( \frac{M_{tomo}}{10^{14} h^{-1} M_\odot} \right)^{0.72 \pm 0.06} h^{-1} M_\odot. \quad (9)$$

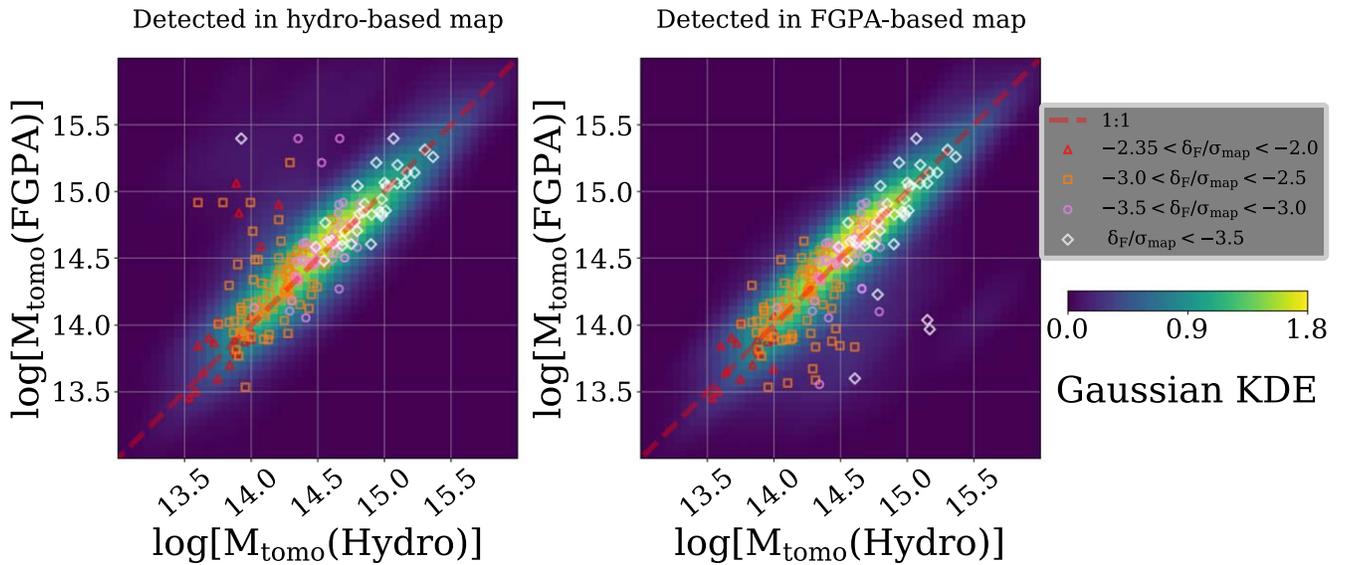

**Figure 10.** The accuracy of the FGPA, applied to a dark-matter-only simulation, for simulating the Lyα flux field is assessed by evaluating the masses of watersheds. The same watershed finder is applied to noiseless maps derived from the full-hydro simulation and from the dark-matter-only simulation using the FGPA. Both panels compare the tomographic masses of watersheds detected in one map to their counterpart watersheds (defined by maximal overlap) in the other. Left panel: comparison of the masses of watersheds detected in the full-hydro map. There is a noticeable set of outliers discussed in the text. Right panel: comparison of the masses of the watersheds detected in the FGPA map. We find that ∼18% of lower-mass watersheds are prone to artificial blending with more massive structures in the FGPA-based map.

The scatter around the mean relation is measured using the remaining four mock observations and is shown on the lower panel of Figure 11. The scatter ranges from 0.35 at the highest masses to 0.55 dex at the lowest. By applying the same method to the noiseless map, we characterize the intrinsic scatter in this relation. The lower panel of Figure 11 shows that this intrinsic error is around 0.30 dex, confirming the scatter is mostly intrinsic and likely originates from the variations in assembly history along with the fairly coarse resolution of the maps, which potentially blends tomographic structures with distinct descendants. Lee et al. (2016) also considered the descendant halos of structures in tomographic maps, and we compare our results in Section 5.

Despite the increasing errors in estimating the descendant masses of the lower-$M_{tomo}$ structures, we emphasize that these watersheds are still preferentially associated with massive $z = 0$ halos. To demonstrate this, we randomly displace the watershed contours within the mock-observed map and determine a new descendant halo mass for each. The distribution of $M_{desc}$ in these randomly located watersheds is shifted to lower masses, as shown in the right panel of Figure 11. For a graphical summary of the procedures and definitions, refer to Figure 12.

### 4.2.2. Progenitors

We now turn to computing the completeness of our method for detecting the progenitors of massive $z = 0$ halos. The method used in Section 4.2.1 is not suitable for this purpose since some watersheds host multiple progenitors of $z = 0$ clusters. For an accurate census of these progenitors, we follow a particle-based approach. We trace back to $z = 2.5$ all of the dark matter particles within a distance $R_{200}$ (where the mean density is $200 \times \rho_{crit}$) of the centers of halos (friends-of-friends groups) at $z = 0$.[10] This method is computationally prohibitive to employ for the many lower-mass halos; therefore, we only consider the progenitors of halos with masses $M_{z=0} > 10^{13.5} h^{-1} M_\odot$, which includes both protoclusters and protogroups. In TNG300-1, there are 1092 halos above this mass threshold. Any progenitor whose center of mass lies within any watershed is considered as "detected." The completeness is then quantified by comparing the total halo mass function at $z = 0$ to the subset with detected progenitors (Figure 13).

The completeness for the smallest structures, protogroups with $10^{13.5} < M_{z=0} / h^{-1} M_\odot < 10^{14.0}$, is ≈20%. However, the completeness increases rapidly with halo mass: more than 40% of protoclusters, $M_{z=0} > 10^{14.0} h^{-1} M_\odot$, are detected, and the completeness for massive protoclusters, $M_{z=0} > 10^{14.5} h^{-1} M_\odot$, is as large as 80%. Figure 13 also shows that the completeness in our analysis is higher than reported by Stark et al. (2015b). This is because we include structures with smaller flux contrast than their threshold of $\delta_F / \sigma_{map} < -3.5$, which increases completeness but also includes more progenitors of lower-mass halos (Figure 11).

Comparing the completeness in the mock and noiseless maps in Figure 13 suggests that observational noise (both from noise in the spectra and finite sightline sampling) cannot entirely account for the incompleteness. To understand the intrinsic difference between the progenitors that are detected and those that are not, we examine their densities in Figure 14. Colors encode the population of the detected and undetected protoclusters in the noiseless map, and density is defined as the spherically averaged dark matter density within $4\ h^{-1}$cMpc of the center of mass. This comparison shows that the tomographic selection is approximately unbiased for protoclusters with masses $M_{z=0} \gtrsim 10^{14.2}\ h^{-1}M_\odot$, while for lower-mass halos, it is biased toward the increasingly small subset of progenitors that were most dense at $z = 2.5$.

---

[10] We neglect gas particles since it is complicated to trace their history using a moving mesh code such as Arepo.

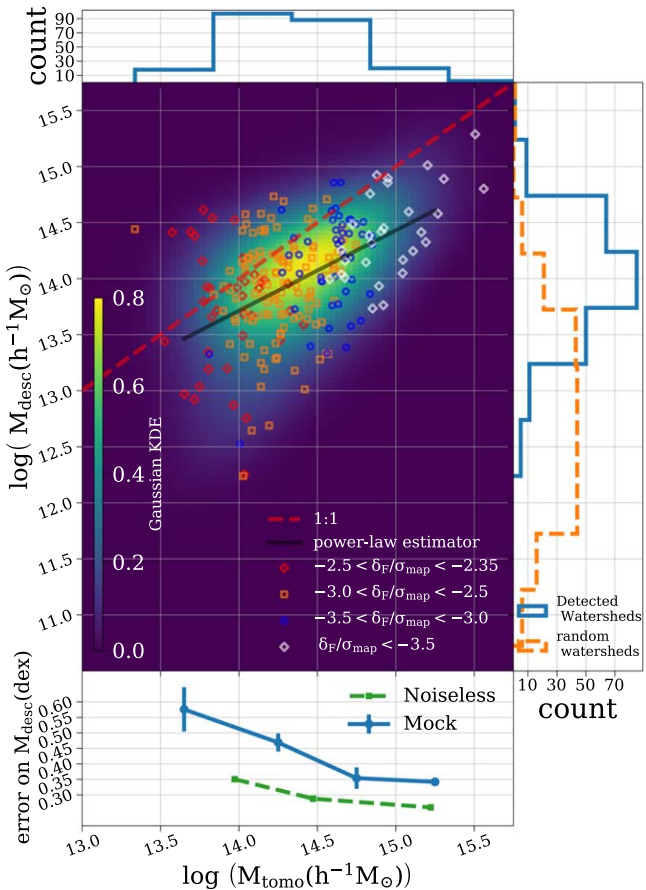

**Figure 11.** The tomographic masses $M_{tomo}$ of watersheds are compared to the masses $M_{desc}$ of their descendant halos at $z = 0$. Main panel: watersheds are colored based on the significance of their peak absorption. The color map illustrates the Gaussian KDE of the data points, and the solid line shows the best-fit power-law estimator $M_{desc}^{est}$ (Equation (9)). Lower panel: the rms error in the estimator $M_{desc}^{est}$, evaluated in several bins of $M_{tomo}$, when $M_{tomo}$ is measured in the mock-observed (solid line) or noiseless (dashed line) map. The comparison of these two curves show that the scatter in this relation is mostly intrinsic (Section 4.2). Top and right panels: the marginal distributions of $M_{tomo}$ and $M_{desc}$. In the right panel, the dashed histogram shows the distribution of $M_{desc}$ when the watersheds are translated to random locations at $z = 0$.

# 5. Summary and Discussion

Tomographically mapping the 3D structure of the Ly$\alpha$ forest with a set of closely spaced sightlines toward background QSOs and LBGs provides a unique opportunity to detect and characterize the extended, massive structures at $z \sim 2.5$ that form today's clusters and groups. Numerical simulations predict these progenitors to be extended over $5$–$10\ h^{-1}$cMpc scales with overdensities of $\delta_m = \rho/\langle\rho\rangle - 1 \approx 2$–$4$ (Chiang et al. 2013). In this work, we build upon the previous intuitive methods developed by Stark et al. (2015b) and Lee et al. (2016) to detect massive structures in Ly$\alpha$ tomographic maps and determine their mass and physical extent. This effort is complicated by significant observational noise in the maps, and by the frequent blending of the progenitors of today's clusters and groups at the resolution of these maps. Compared to earlier methods, we aim to include lower-mass structures while mitigating the associated increase of noise and blending. Below, we briefly summarize our method and its two tunable parameters, which we optimized for an LATIS-like survey at $z = 2.5$ by testing against a large hydrodynamical simulation. A graphical summary is presented in Figure 12.

*Detecting structures:*

1. Draw contours of $\delta_F/\sigma_{map} \leqslant \nu = -2$ to define "islands" within the $\delta_F$ map, which is smoothed by a Gaussian kernel with $\sigma_{sm} = 4\ h^{-1}$cMpc (Section 3.1).

2. Within each island, identify all local flux minima with $\delta_F/\sigma_{map} \leqslant \kappa = -2.35$. If no such minima exist, remove that island. If multiple such minima are present, use the watershed algorithm to partition the island into watersheds associated with each minimum (Section 3.2 and Figure 4).

*Estimating structures' masses at $z = 2.5$:*

1. Use the second-degree polynomial fit in Equation (5) to relate the measured $\delta_F$ to the dark matter density within each voxel of a watershed (Section 2.4.2 and Figure 2).

2. Integrate these DM densities over the watershed's volume and apply the offset factor in Equation (7) to estimate a tomographic mass $M_{tomo}$ (Section 4.1 and Figure 8).

3. Use Equation (8) to estimate the DM mass of the watershed from $M_{tomo}$. This estimator corrects for the bias from the noisy boundary. Uncertainties are provided in Section 4.1. (Figure 8).

*Estimating structures' descendant masses at $z = 0$:*

1. Use the measured $M_{tomo}$ and the power law in Equation (9) to estimate $M_{desc}$, the mass of the descendant halo at $z = 0$. Uncertainties are shown in Figure 11.

This method can be applied to LATIS-like tomographic maps to produce catalogs of structures (watersheds) that range from protogroups ($M_{z=0} = 10^{13.5}$–$10^{14}\ h^{-1}M_\odot$) to protoclusters ($M_{z=0} > 10^{14}\ h^{-1}M_\odot$). We estimate the method's completeness is 80% for the most massive protoclusters (Figure 13). Less massive structures leave smaller signals within the absorption map; however, our method still achieves a 20% completeness for protogroups and is biased toward those embedded within denser regions on scales of $4\ h^{-1}$cMpc (Figure 14).

We find the scatter in the $\delta_F^{sm}$–$\rho_{DM}^{sm}$ relation results in a remarkably small scatter of 0.12 dex (left panel of Figure 8) in the estimated mass of structures at $z \sim 2.5$, if the mass is measured within a fixed aperture, such as an observed flux contour. This result is similar to that shown in Figure 9 of Lee et al. (2016). There are also Bayesian methods designed to infer the underlying DM density and velocity by comparing the approximate $N$-body simulations with tomography observations (Horowitz et al. 2019, 2021a, 2021b). They showed by adding information from the coeval galaxy distribution to the Ly$\alpha$ tomography, their method reduces the scatter in protocluster mass estimation by a factor of two if one defines the mass as an integrated spherical overdensity within a radius around the Eulerian position of the corresponding cluster at $z = 0$ (i.e., comparing the left panel of Figure 8 and that of Figure 10 in Horowitz et al. 2021b). However, we find that the dominant uncertainty in the mass estimate is due to the uncertainty in measuring the structures' volumes. When comparing each mock-observed structure to the overlapping structure in the noiseless map, the uncertainty in the mass estimate increases (right panel of Figure 8), particularly for lower-mass structures, and $M_{tomo}$ becomes a biased estimator. We provide an unbiased estimator $M_{DM}^{est}$ and find that its uncertainty ranges from 0.29 to 0.36 dex for the highest to lowest mass bins.

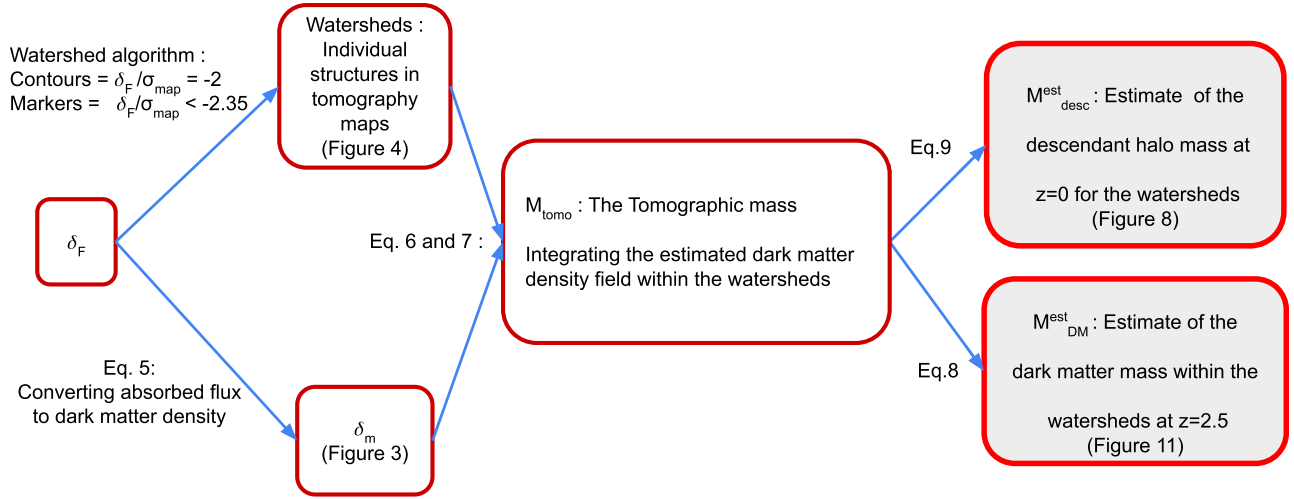

**Figure 12.** A step-by-step guide to our method for detecting structures in a Lyα tomographic map, estimating the dark matter mass within them at $z = 2.5$, and estimating their descendant halo masses at $z = 0$.

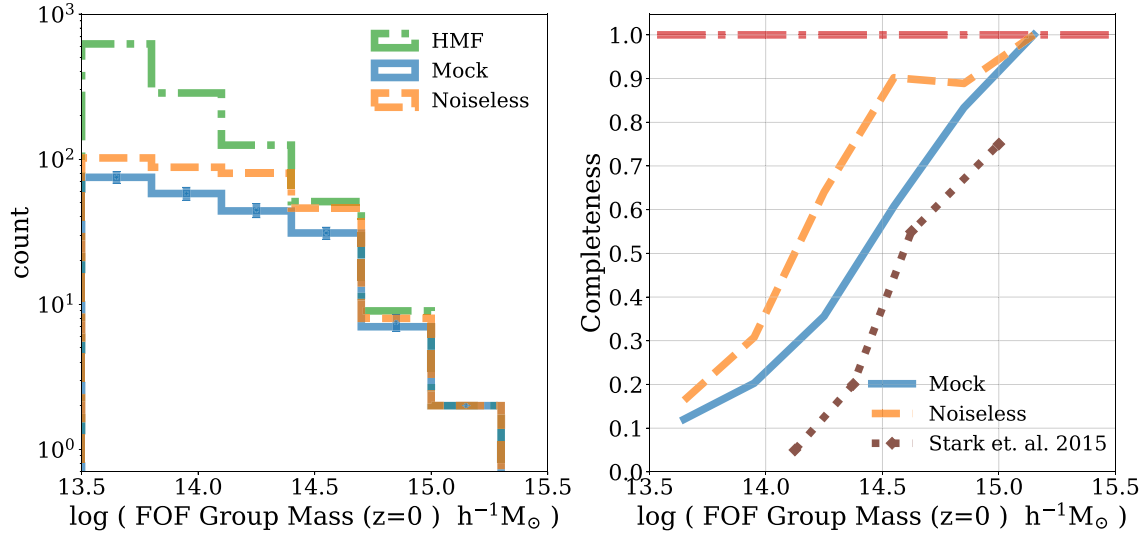

**Figure 13.** Left panel: the halo mass function at $z = 0$ (dotted–dashed curve) is compared to the mass distribution of the halos whose progenitors are detected in the mock-observed (solid curve) and noiseless maps (dashed curve). For the mock-observed maps, the median distribution among 20 mock maps and the $1\sigma$ scatter are shown. Right panel: the completeness, defined as the fraction of massive $z = 0$ halos whose progenitors are detected in the mock-observed (solid curve) or noiseless (dashed curve) maps. Our method detects progenitors of halos over about one decade of mass, from massive protogroups to protoclusters. The completeness reported by Stark et al. (2015b; dotted curve) is slightly lower due to their choice of a different detection threshold.

This precision is competitive with the estimates from dynamical methods applied to galactic spectroscopic surveys. For instance, Cucciati et al. (2018) estimated a median error of 0.35 dex for masses of components of the Hyperion proto-supercluster, which have a median mass of $10^{13.8} h^{-1} M_\odot$. Similarly, Darvish et al. (2020) spectroscopically confirmed a protocluster in the COSMOS field at $z \sim 2.2$ and estimated the dynamical mass to be $1–2 \times 10^{14} h^{-1} M_\odot$. On the other hand, dynamical mass estimates are strictly applicable only to virialized systems, which is not the case for protoclusters or protogroups. Lyα tomographic mass estimates do not rely on this assumption. The other method commonly used to estimate the masses of structures at high redshifts is the galactic overdensity (e.g., Steidel et al. 1998). More recently,

Ata et al. (2021) developed a Bayesian method to reconstruct the underlying DM density field for a combination of multiple independent spectroscopic galactic surveys in the COSMOS field. However, all methods based on galactic surveys require assuming that all of the observed galactic subpopulations trace the density field according to a simple model (i.e., same bias factor). Another key benefit of Lyα tomographic masses is their independence from the galactic content of a structure, which is valuable for galaxy–environment studies.

We also estimated the mass at $z = 0$ of the descendants of structures detected in tomographic maps (Section 4.2). These estimates are significantly more uncertain, about 0.4 dex (Figure 11). We have verified the scatter is only slightly reduced in the noiseless map, which suggests the uncertainty is driven mostly by the scatter in assembly history, as concluded by Stark et al. (2015b). Lee et al. (2016) also considered the descendant halos of structures detected in tomographic maps. We find two significant differences with their results. The slope of the $M_{\rm tomo}$–$M_{\rm desc}$ relation (Equation (9)) is closer to unity in our analysis (0.72 versus 0.27), and our estimate of the scatter is larger (0.3–0.45 dex versus ~0.2 dex). There are several explanations for these differences. First, when defining this relation, Lee et al. (2016) considered only the structures in the mock-observed map that are actually progenitors of cluster-scale halos. We find that censoring the 25% of structures with lower-mass descendants suppresses the slope and the scatter. Second, Lee et al. (2016) chose a different threshold ($\delta_F / \sigma_{\rm map} < -3$) to define the integration contours, which encloses a different fraction of the collapsed halo mass. Third, we impose a margin between the integration contour and the minimum $\delta_F$ within a watershed (i.e., $\kappa < \nu$), which effectively places a lower bound on $M_{\rm tomo}$ by eliminating very small contours. We tested the sensitivity of structure detection and mass estimates in mock tomographic maps to the detailed modeling of the gas and galactic evolution (Section 4.1.1). We conclude that the FGPA applied to dark-matter-only simulations imputes sizeable errors in characterizing lower-mass structures, $M_{\rm tomo} \lesssim 10^{14.2} \, h^{-1} M_\odot$. Compared to full-hydro simulations, FGPA confuses lower-mass neighboring structures by either merging them with more massive neighbors or fracturing them into few smaller watersheds. However, contrary to Kooistra et al. (2022b), we find the global distribution of $\delta_F^{\rm sm}$–$\rho_{\rm DM}^{\rm sm}$ relation (Figure 9) at the relevant scales ($4 \, h^{-1}$ cMpc) and redshifts ($z \sim 2.5$) is insensitive to the galactic feedback and detailed hydro prescription in TNG simulations. The IllustrisTNG simulations also do not include all of the physics that may be relevant. Although the galactic formation model approximately accounts for the local heating due to black hole accretion (Section 2.6.4 in Vogelsberger et al. 2013), to fully account for the proximity effect, the local ionization should be modeled. This requires radiative transfer simulations, which are computationally demanding. However, Miller et al. (2021) placed an approximate bound on this effect by sampling the AGN luminosity function and assigning AGNs to massive halos in a one-to-one rank order fashion (implying a duty cycle of one). They conclude local ionization usually has a small effect on the Ly$\alpha$ absorption observed in tomographic maps, even within massive protoclusters.

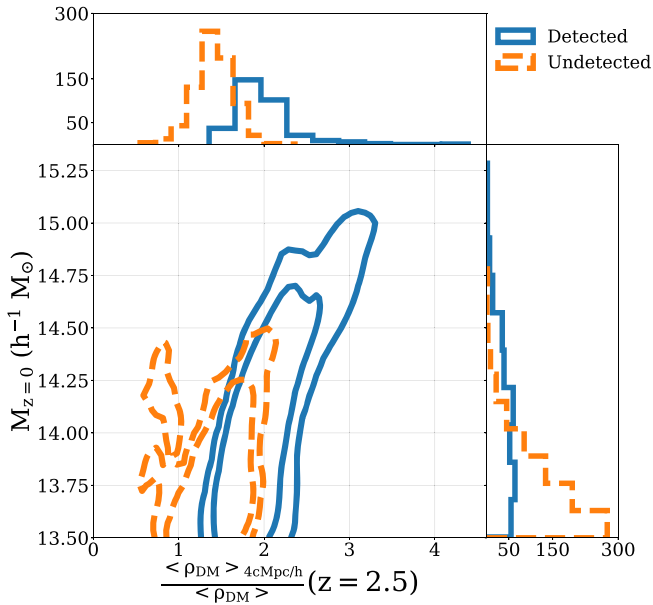

**Figure 14.** Comparison of the densities of the progenitors of massive $z = 0$ halos ($M_{z=0} > 10^{13.5} \, h^{-1} M_\odot$) that are detected (solid curves) or undetected (dashed curves) in the noiseless $z = 2.5$ flux map. The x-axis shows the ratio of the average dark matter density, measured within a sphere of radius $4 \, h^{-1}$ cMpc around the center of mass of each progenitor, to the global mean dark matter density at $z = 2.5$. The detected progenitors of massive halos are an approximately unbiased subset, but for lower-mass halos, the detection is biased toward the progenitors embedded in denser environments at $z = 2.5$. For each population, the contours show the 68% and 5% levels of the 2D Gaussian KDE.

Moreover, X-ray observations of galaxy clusters provide evidence for some feedback activities (e.g., Edge & Stewart 1991; Markevitch 1998). It is speculated that a form of "preheating" started early on in the progenitors of these clusters (e.g., Evrard & Henry 1991). Kooistra et al. (2022a) studied this effect in simulated tomography maps by implementing simple feedback models, i.e., injecting energy or entropy to all particles within protoclusters, using the broad constraints on the feedback strength from Chaudhuri et al. (2013) and Iqbal et al. (2017). They predicted significant ionization within protoclusters in cases with strong feedback. However, we do not find increased transparency in the overdense regions due to feedback when comparing the FGPA and full-hydro noiseless maps (Figure 9). The feedback prescriptions in the TNG300-1 simulation are state of the art (Section 2.1 and Weinberger et al. 2017). Additionally, our mock-observed tomography maps match the observed flux distribution in LATIS very well (Figure 3), while much stronger feedback could disturb the agreement for $\delta_F < 0$.

The methods developed in this work for detection and characterization of the structures in an LATIS-like tomography survey, as well as the realistic mock maps built here, will be used in future works. The catalogs of structures and their masses derived with these methods will be essential ingredients in studying the environmental dependencies of galaxy properties, e.g., star formation and metallicity, within statistically representative samples of overdensities at "Cosmic Noon." The two parameters in our method can be tuned for any other Ly$\alpha$ tomography survey with different spectroscopic noise properties or sightline density.

M.Q. was supported by a UCR-Carnegie Graduate Fellowship and NSF grant AST-2107821. S.B. was supported by NSF grant AST-1817256. A.B.N. acknowledges support from the National Science Foundation under grant AST-2108014. Computing resources were provided by NSF XSEDE allocation AST180058 and high performance computing centers at Carnegie institute and UCR. This work made use of the open source python package Astropy.

### Data Availability

Illustris-TNG is publicly available. Our analysis scripts and notebooks on how to use the codes are available at 10.5281/zenodo.6365915. All produced data can be downloaded from 10.5281/zenodo.5770882.

### Appendix A
### 1D Flux Power Spectrum

Figure 15 compares the noiseless 1D flux (i.e., $\delta_F$) power spectrum simulated with the full-hydro TNG300-1 and the FGPA models to the observational constraints from SDSS DR14 (Chabanier et al. 2019). We estimate the power spectra by applying the fast Fourier transform on ∼8000 simulated spectra and correct for the finite pixel resolution effect. The convergence test is passed with half as many spectra. The oscillations in the observed power spectrum arising from Lyα–SiIII correlations are not modeled in the simulated spectra. After correcting for that, we expect the simulated power spectra in both methods to be consistent with observations with 10% accuracy.

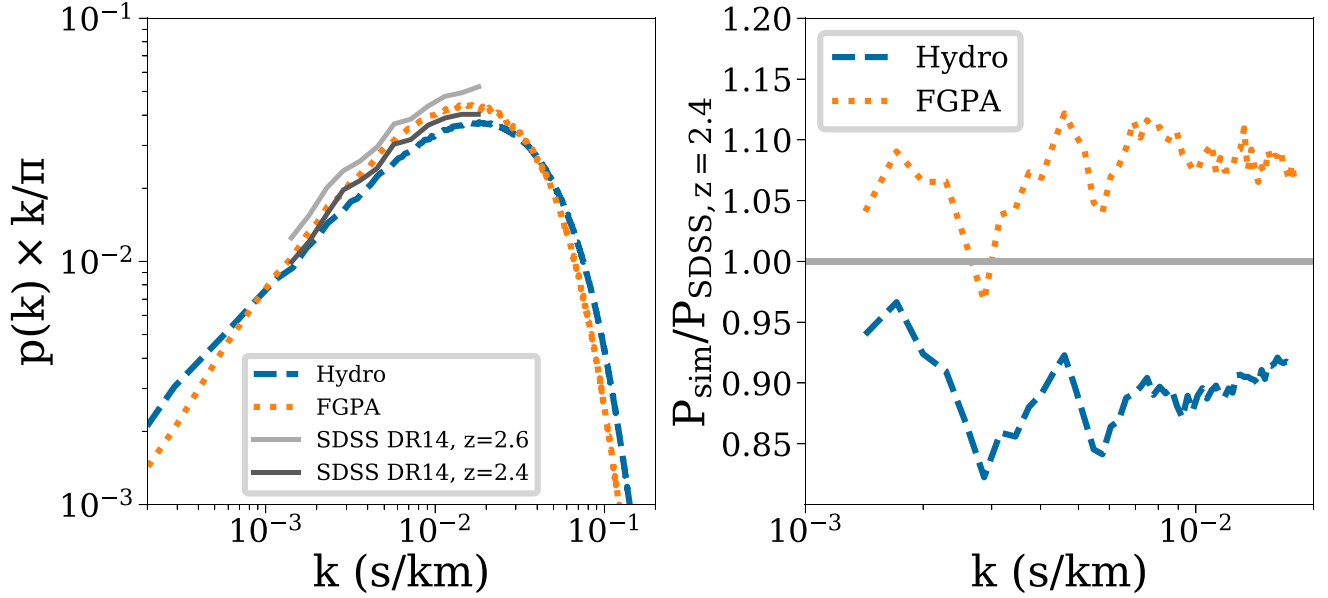

**Figure 15.** The noiseless 1D flux power spectra simulated with both the full-hydro and FGPA methods are compared to the observations in SDSS DR14 (Chabanier et al. [2019]). In the right panel, the ratio of the simulated power spectra to observations shows a ∼10% agreement for both methods.

## Appendix B
## Sensitivity of FGPA Accuracy to Smoothing Scale

In this section we further examine the differences between the approximate FGPA approach and the full-hydro simulations in detail. For a lateral comparison with the recent findings in Kooistra et al. ([2022b](#)), we extract the 3D noiseless mock absorbed flux map from the TNG100-1 simulation at $z = 2.0$ using both hydro and FGPA prescriptions. The voxel resolution of each map is set to $V_{voxel} = 125 \, h^{-1}$ckpc. On the left panel in Figure [16](#), the maps are smoothed with a kernel size of $\sigma_{sm} = 3 \, h^{-1}$cMpc, similar to Kooistra et al. ([2022b](#)). Contrary to Kooistra et al. ([2022b](#)), our comparison reveals more similarity between the two methods (compare the solid and dashed contours). The dotted contour, however, illustrates that one could eliminate this small difference by adjusting the initial smoothing scale on the DM density and velocity in the FGPA method, set to mimic baryonic pressure smoothing, to the value suggested by previous works (e.g., Sorini et al. [2016](#)), i.e., $\lambda_{sm} \sim 250 \, h^{-1}$ckpc. In the right panel, we instead smooth the mock maps with our fiducial kernel, i.e., $\sigma_{sm} = 4 \, h^{-1}$cMpc. Comparison between the two panels shows that a slightly larger smoothing scale erases the small scale disparity between the two methods. This is in agreement with our conclusion in Section [4.1.1](#) that using the FGPA method to generate maps smoothed with $\sigma_{sm} = 4 \, h^{-1}$cMpc affects only a subdominant fraction of detected progenitors (∼18%).

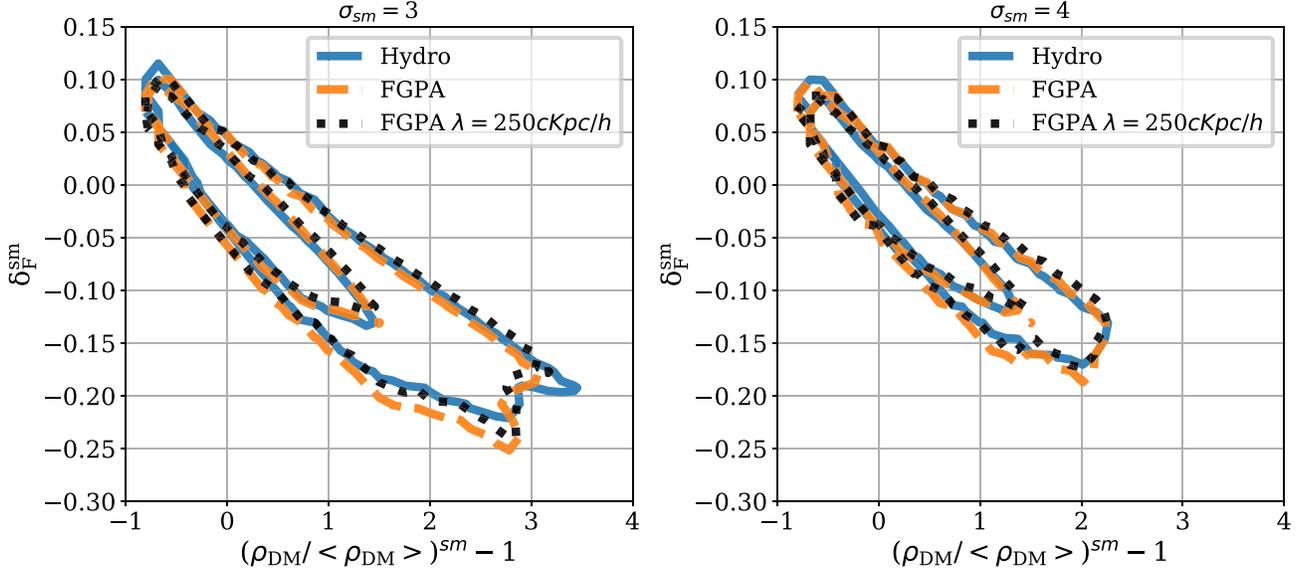

**Figure 16.** Similar to Figure 9, we compare the approximate FGPA method applied to DM-only simulations with spectra from a full-hydro simulation. The noiseless mock maps are generated from TNG100-1 simulations at $z = 2.0$ and are smoothed with different kernel sizes of $\sigma_{sm} = 3, 4$ $h^{-1}$cMpc on the left and right panels, respectively. The FGPA map is increasingly similar to hydro as the smoothing scale is increased. The dotted line represents the FGPA map generated after an initial smoothing of the DM density and velocity to mimic baryonic pressure smoothing.

## Appendix C
## Redshift Sensitivity

In an Ly$\alpha$ tomographic survey that maps a significant redshift interval (for LATIS, $z = 2.2$–2.8), there will be both cosmological evolution (e.g., in the mean density and mean flux) as well as a redshift dependence in the observational parameters (e.g., the sightline density and the noise in the observed spectra). Throughout this work, we fixed these parameters to those at the simulation snapshot $z = 2.45$, near the redshift midpoint of LATIS. To test the accuracy of our mass estimators at different redshifts within the LATIS map, we built 20 mock-observed maps at the $z = 2.3$ and $z = 2.6$ simulation snapshots using the LATIS sightline density and noise (Table 1) at the corresponding redshifts. This translates to 9017 and 4409 spectra for redshifts $z = 2.3$ and $z = 2.6$, respectively, compared to 6739 at $z = 2.5$. The estimator of the $\delta_F^{sm}$–$\rho_{DM}^{sm}$ relation in Equation (5) changes slightly within this redshift range, as shown in Table 2. However, these slight changes result in a negligible offset, $\sim 0.02$ dex, in the tomographic mass measurements $M_{tomo}$. After fixing the $\delta_F^{sm}$–$\rho_{DM}^{sm}$ estimator to $z = 2.5$ and adopting the same watershed parameters for defining structures, i.e., $\nu = -2$ and $\kappa = -2.35$, we find that the best-fit parameters in the $M_{DM}^{est}$–$M_{tomo}$ and $M_{desc}^{est}$–$M_{tomo}$ estimators are consistent within this redshift range, as shown in Tables 3 and 4. We conclude that the estimators we derived at $z = 2.5$ can be applied throughout the LATIS map volume.

**Table 2**
Redshift Evolution in the $\delta_F^{sm}$–$\rho_{DM}^{sm}$

Relation $\left(\frac{\rho_{DM}}{\langle\rho_{DM}\rangle}\right)^{sm} = a_2\,\delta_F^{sm\ 2} + a_1\,\delta_F^{sm} + a_0$

| z | $a_2$ | $a_1$ | $a_0$ |
|---|---|---|---|
| $z = 2.3$ | $14.6 \pm 0.6$ | $-4.97 \pm 0.06$ | $0.97 \pm 0.01$ |
| $z = 2.5$ | $16.8 \pm 0.6$ | $-5.62 \pm 0.07$ | $0.96 \pm 0.01$ |
| $z = 2.6$ | $15.23 \pm 0.7$ | $-5.44 \pm 0.06$ | $0.98 \pm 0.01$ |

**Table 3**
Redshift Evolution in $M_{DM}^{est} = a \times \left(\frac{M_{tomo}}{10^{14}}\right)^b$ in Units of $h^{-1}M_\odot$

| z | a | b |
|---|---|---|
| $z = 2.3$ | $10^{14.59 \pm 0.06}$ | $0.43 \pm 0.09$ |
| $z = 2.5$ | $10^{14.54 \pm 0.04}$ | $0.39 \pm 0.07$ |
| $z = 2.6$ | $10^{14.63 \pm 0.04}$ | $0.35 \pm 0.05$ |

**Table 4**
Redshift Evolution in $M_{desc}^{est} = a \times \left(\frac{M_{tomo}}{10^{14}}\right)^b$ in Units of $h^{-1}M_\odot$

| z | a | b |
|---|---|---|
| $z = 2.3$ | $10^{13.69 \pm 0.06}$ | $0.76 \pm 0.09$ |
| $z = 2.5$ | $10^{13.71 \pm 0.06}$ | $0.72 \pm 0.09$ |
| $z = 2.6$ | $10^{13.70 \pm 0.05}$ | $0.77 \pm 0.09$ |

## Appendix D
## Other Smoothing Scales

Stark et al. (2015b) showed that smoothing the flux absorption map with a $\sigma_{sm} = 4\ h^{-1}$cMpc Gaussian kernel behaves as a matching filter for detecting protoclusters, and our analysis throughout the paper is based on maps smoothed in this way. Since one goal of this paper is to expand the detections to smaller structures like protogroups, i.e., with $M_{z=0} < 10^{14}\ h^{-1}M_\odot$, we test here whether smaller kernels are a better choice for detecting protogroups. Specifically, we investigate the watersheds that are only found in less-smoothed maps and do not overlap with any watershed in maps smoothed with the standard kernel of $\sigma_{sm} = 4\ h^{-1}$cMpc.

We follow the same steps discussed in Sections 3.1 and 3.2 on the maps smoothed with a smaller $\sigma_{sm} = 2\ h^{-1}$cMpc kernel, and find that the optimal parameters are $\nu = -2.30$ and $\kappa = -2.75$. Proceeding with the steps in Section 4.2, except using the new parameters, we derive the $M_{desc}$–$M_{tomo}$ relation for the detected watersheds, which is shown in Figure 17. The orange stars indicate the watersheds that do not overlap with any of the watersheds within the map smoothed with $\sigma_{sm} = 4\ h^{-1}$cMpc; these constitute 16% of the structures. The rest are colored based on their absorption significance. On one hand, the marginal histograms over $M_{desc}$ (right panel) show that a fraction of the new structures found in the $\sigma_{sm} = 2\ h^{-1}$ cMpc map are indeed significant overdensities, some associated with massive protoclusters. On the other hand, in comparison to the results for $\sigma_{sm} = 4\ h^{-1}$cMpc map, the scatter in the estimated descendant mass is much larger. Figure 18 shows the rms error, similar to bottom panel of Figure 11, but plotted versus $M_{desc}^{est}$ since the $M_{tomo}$ measured in maps smoothed differently are not directly comparable. Due to the larger noise in less-smoothed maps, the uncertainty in $M_{desc}$ is larger at every mass scale. We conclude that maps with lower smoothing may be used to identify some lower-mass structures, but at the cost of larger noise contamination.

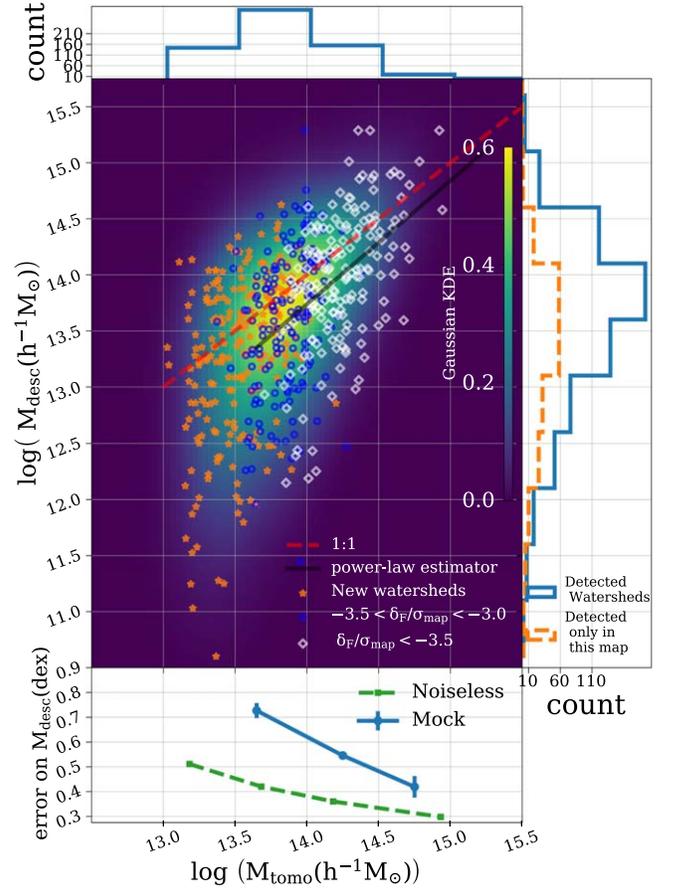

**Figure 17.** Same as Figure 11, but for the mock map smoothed with a smaller $\sigma_{sm} = 2\ h^{-1}$cMpc Gaussian. Red stars indicate the new watersheds detected only in this less-smoothed map. The other watersheds are colored based on their peak absorption significance.

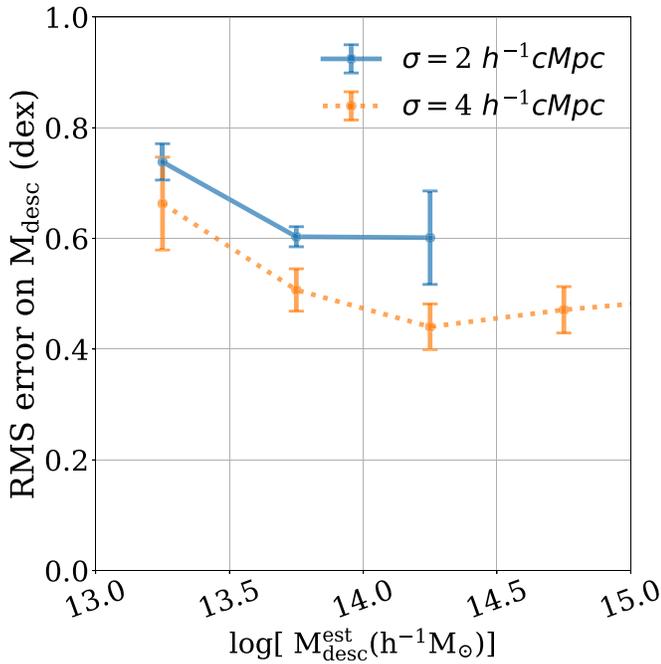

**Figure 18.** Comparing the uncertainty in descendant mass estimation between the mock-observed maps smoothed with $\sigma = 2\ h^{-1}$cMpc and $\sigma = 4\ h^{-1}$cMpc kernels. The *x*-axis shows the descendant mass $M_{\rm desc}^{\rm est}$ estimated via the power-law estimator for each smoothing scale (Figures 11 and 17). At all masses, the precision of the prediction of the descendant mass is worse when using the less-smoothed map.

**ORCID iDs**

Mahdi Qezlou ● https://orcid.org/0000-0001-7066-1240
Andrew B. Newman ● https://orcid.org/0000-0001-7769-8660
Gwen C. Rudie ● https://orcid.org/0000-0002-8459-5413
Simeon Bird ● https://orcid.org/0000-0001-5803-5490